\documentclass[journal]{IEEEtran}

\usepackage{cite}
\usepackage[printonlyused,nolist]{acronym}
\usepackage[colorinlistoftodos,textwidth=10mm,disable]{todonotes}
\usepackage{float}
\usepackage{mathtools}
\usepackage{amssymb}
\usepackage{url}
\usepackage{datetime}
\usepackage{booktabs}
\usepackage{changes}
\usepackage[utf8]{inputenc}
\definechangesauthor[color=red]{jonas}
\definechangesauthor[color=purple]{aidin}
\definechangesauthor[color=blue]{anders}

\ifCLASSINFOpdf
  \usepackage{graphicx}
\else
\fi

\usepackage{bm}
\usepackage{array}

\begin{document}

\title{Angular sampling, Test Signal, and Near Field Aspects for Over-the-Air Total Radiated Power Assessment in Anechoic Chambers}

\author{Jonas~Frid\'en, Aidin Razavi, Anders Stjernman
\thanks{Ericsson Research, Ericsson AB}
}

\markboth{Article submitted to a journal}%
{Fridén, Razavi, Stjernman \today}

\maketitle

\begin{abstract}
\acp{5G} operating in \ac{MMW} bands will employ
base stations with integrated \acp{AAS} capable of beam-tracking using narrow beams obtained from array antennas encapsulated in a chip-like device. Regulatory limits for unwanted \ac{RF} emissions are currently set in terms of \ac{TRP}. As measurements at the antenna connectors are not possible, \ac{OTA} methods for  \ac{TRP} of unwanted emissions are needed. The method investigated here uses power density measurements on a spherical surface in an anechoic chamber. Two major challenges with such a method are: need for large number of angular points, and search for worst case antenna configuration per frequency for electrically large devices at high frequencies. These challenges are addressed by investigating the impact of correlation, sparse sampling, and use of beam sweeping on the \ac{TRP} estimate. Finally, it is investigated how and in which spatial regions near-field tangential electric field measurements can be used to assess \ac{TRP}. 
\end{abstract}

\begin{acronym}
\acro{AC}{Anechoic Chamber}
\acro{AAS}{Active Antenna System}
\acro{CATR}{Compact Antenna Test Range}
\acro{CDF}{Cumulative Probability Distribution}
\acro{FEM}{Finite Element Method}
\acro{EIRP}{Equivalent Isotropically Radiated Power}
\acro{EMC}{Electromagnetic Compatibility}
\acro{EUT}{Equipment Under Test}
\acrodefplural{EUT}{Equipments Under Test}
\acro{FCC}{Federal Communications Commission}
\acro{HPBW}{Half Power Beam Width}
\acroindefinite{HPBW}{an}{a}
\acro{MA}{Measurement Antenna}
\acroindefinite{MA}{an}{a}
\acro{MMW}[mmW]{milli-meter Wave}
\acroindefinite{MMW}{an}{a}
\acro{MU-MIMO}{Multi User Multiple-Input-Mutiple-Output}
\acroindefinite{MU-MIMO}{an}{a}
\acro{OTA}{Over-The-Air}
\acro{PM}{Pattern Multiplication}
\acro{RBW}{Resolution Band Width}
\acroindefinite{RBW}{an}{a}
\acro{RWG}{Rectangular Wave Guide}
\acroindefinite{RWG}{an}{a}
\acro{RC}{Reverberation Chamber}
\acroindefinite{RC}{an}{a}
\acro{RF}{Radio Frequency}
\acro{SF}{Sparsity Factor}
\acroindefinite{SF}{an}{a}
\acro{SGH}{Standard Gain Horn}
\acroindefinite{SGH}{an}{a}
\acro{SNR}{Signal-to-Noise Ratio}
\acroindefinite{SNR}{an}{a}
\acro{SSCS}{Standard Spherical Coordinate System}
\acroindefinite{SSCS}{an}{a}
\acro{SWE}{Spherical Wave Expansion}
\acroindefinite{SWE}{an}{a}
\acro{TRP}{Total Radiated Power}
\acro{3GPP}{Third Generation Partnership Project}
\acro{5G}[5G system]{5th Generation Mobile Network System}
\end{acronym}


\IEEEpeerreviewmaketitle

\def\ie{\emph{i.e.}}
\def\eg{\emph{e.g.}}
\def\viz{\emph{viz.}}
\def\ff{far-field}
\def\nf{near-field}
\def\FF{Far-field}
\def\NF{Near-field}
\renewcommand\vec[1]{\mbox{\boldmath $#1$}}
\newcommand\mat[1]{\mathbf{#1}}
\newcommand\colvec[1]{[\mathbf{#1}]}
\newcommand\unitvec[1]{\boldsymbol{\hat{\textbf{$#1$}}}}
\newcommand\partder[2]{\frac{\partial #1}{\partial #2}}
\newcommand\partderInline[2]{\partial #1/\partial #2}
\def\ju{\mathrm{j}}
\def\eu{\mathrm{e}}
\def\Re{\mathrm{Re}}
\def\Im{\mathrm{Im}}
\def\dOmega{\sin\theta\mathrm{d}\theta\mathrm{d}\phi}
\def\TRP{\ensuremath{\mathrm{TRP}}}
\def\EIRP{\ensuremath{\mathrm{EIRP}}}
\newcommand\fsav[1]{\left\langle #1\right\rangle_\mathrm{fs}}
\newcommand\gridav[1]{\left\langle#1\right\rangle_\mathrm{grid}}
\def\SF{\ensuremath{\mathrm{SF}}}
\def\hpbw{\ensuremath{\mathrm{HPBW}}}
\def\rmeas{\ensuremath{r_0}}

\def\Rsph{R_\mathrm{sph}}
\def\Rcyl{R_\mathrm{cyl}}
\def\Dsph{D_\mathrm{sph}}
\def\Dcyl{D_\mathrm{cyl}}

\DeclarePairedDelimiter\ceil{\lceil}{\rceil}
\DeclarePairedDelimiter\floor{\lfloor}{\rfloor}


\section{Introduction}

\acp{5G}~\cite{3GPPTS38.104} are envisioned to make use of \ac{MU-MIMO} and advanced beam forming techniques enabled by tightly integrated active antenna arrays implying that \ac{RF} emissions cannot be measured at the antenna port. The \ac{FCC} has already released the technical conditions for using \acp{MMW} adopting a novel approach for setting the unwanted emission limits in terms of \ac{TRP}, instead of traditional \ac{EIRP} limits~\cite[p.~106]{FCC-16-89A1}. A similar approach is agreed in \ac{3GPP} standardization~\cite[sec.~9.7]{3GPPTS38.104}. At this moment, a standardized methodology for evaluating the \ac{TRP} of the unwanted emissions is not yet established. Some challenges presented by a method based on integration of far-field radial power flux density around the \ac{EUT} placed in a full anechoic chamber are investigated. The main practical limitations are: long test time and large test distance. Note, there is no theoretical uncertainty or ambiguity when it comes to defining and calculating the \ac{TRP} of a radiating device.

For lower frequencies and electrically small antennas the \ac{RC} is a well established method \cite{IEC2011,Kildal2012,Hill1998} for \ac{TRP}-measurements. However, at higher frequencies the current status and availability at test houses is limited. Therefore, the scope of this paper is \acp{CATR} and \acp{AC}~\cite{CTIA2016}.

Total test time to prove compliance to the regulatory limits is of major concern to regulators, manufacturers and test labs. Test time is directly proportional to the number of necessary test points in the frequency and spatial domain and the total number of device configurations to be tested.

According to present regulations,  the domain of spurious emissions for \ac{MMW} devices ranges from a few tens of MHz up to 100 GHz or more~\cite{FCC-16-89A1,3GPPTS38.104}. If all this spectrum needs to be investigated by means of measurements on dense enough spherical grids for every single MHz, the test time will become very large compared to today's test time for similar products operating under 6 GHz. An important step here, is to identify an efficient and practical way of finding the frequencies of potentially high emissions requiring further investigation, \emph{a.k.a.} pre-scan~\cite{ANSIC63.26-2015}. An important matter albeit it is not treated here and it is assumed that the relevant frequencies can be found.

A key parameter for the test time for a single frequency is the angular steps needed to assess the \ac{TRP} with negligible error. The corresponding set of test points is hereafter denoted a \emph{dense sphere}. The underlying theory is well known \cite{Hald1998} and the number of needed test points depends on the size of the \ac{EUT} and the frequency. However, this leads to an impractical measurement time for electrically large devices and emission characterization at high frequencies. To reach an acceptable measurement time the effect on the error of \ac{TRP} when using sparser angular sampling, than in the \emph{dense sphere} case, is investigated. To adopt to simpler, but still common, measurement ranges and turntable layouts also measurements in a few cuts are considered. This introduces a risk of estimating an erroneous value of \ac{TRP} which is handled by adding a margin to the \ac{TRP} value assessed directly from the grid. This margin varies with the sparsity of the grid and the grid type and provides a means for trading measurement accuracy against measurement time. Low emissions can be measured with fewer test points since a higher margin can be used, while emissions with \ac{TRP} close to the limit will require a lower margin and hence a denser grid implying longer measurement time.

For large antenna arrays with individual phase and amplitude control per antenna element, varying carrier band width, and modulation format, \eg\ \ac{MU-MIMO} systems, the number of possible antenna configurations can be huge. Current regulations indicate the need to find the worst case, which might lead to a lengthy exhaustive search through all states of the antenna system, having in mind that the worst case configuration may change with frequency. To avoid investigation of narrow beam patterns at different antenna configurations a solution based on beam sweeping during the test time is probably a good compromise. By sweeping through all possible, or a selected group, of beam positions and measuring the average pattern, the emission peaks will be broadened and a more sparse measurement grid can be used to characterize the spatial distribution of the radiated power density. Hence, beam sweeping has a potential to reduce the measurement time both by reducing the needed number of angular points, and by reducing the number of configurations to test.

Traditional \ac{EMC} testing based on \ff\ parameters, such as the \ac{EIRP}, require measurements at distances larger than the \ff\ distance $\approx 2D^2/\lambda$, where $D$ is the diameter of the \ac{EUT} and $\lambda$ is the wavelength. The \ff\ distance for a \ac{MMW} base station can be more than 10 meters, even hundred meters in some cases, which is unpractical. To avoid \ff\ test distances, radial power flux density in the \nf\ is addressed. 

The topics addressed in the following sections are: 1) trade-off between accuracy of the \ac{TRP} estimation and the number of angular points, 2) effect of beam sweeping on test time and uncertainty, and 3) how and where to do measurements in the \nf\ region to get an accurate estimation of the \ac{TRP}. The conclusions of this paper can serve as a guidance to regulatory approval of upcoming \ac{MMW} devices.

\section{Definitions and basic relations} \label{sec:basics}

The \acf{TRP} is by definition the \emph{power radiated by the antenna}~\cite[Figure 1]{IEEE145-2013}. The term \emph{total} is used to emphasize that the sum of  the partial contributions from a complete set of polarizations is to be used. Moreover, it is a function of frequency and antenna configuration. From a measurement point of view the \ac{TRP} is the sum of all power flowing out of the antenna through a measurement surface that captures all the outgoing power. If the antenna port is accessible the measurement surface can be selected as the cross section of the connected cable at the antenna port. If the antenna port is not accessible, \ie\ for integrated antenna systems~\cite{IEEE145-2013}, the measurement surface needs to enclose the entire \ac{EUT}, which leads to \ac{OTA} testing. In this case \ac{TRP} can be obtained by measuring the radial power flux density at the measurement surface, unit $\mathrm{W/m^2}$, and integrate these values over the measurement surface. The magnitude and direction of the power flux density of the electromagnetic field is given by the Poynting vector. In the case of time-harmonic fields, $\vec{E}(\vec{r},t) = \Re[\sqrt{2}\vec{E}(\vec{r})\exp{(\ju \omega t)}]$ and $\vec{H}(\vec{r},t) = \Re[\sqrt{2}\vec{H}(\vec{r})\exp{(\ju \omega t)}]$, the time average of the Poynting vector is~\cite{Cheng1989}

\begin{equation}\label{eq:poyntingVector}
\vec{S}(\vec{r}) = \Re[\vec{E}(\vec{r})\times\vec{H}(\vec{r})^*].
\end{equation}
Here,  effective values of the fields are used. In this investigation only spherical measurement surfaces will be considered. The \ac{SSCS}~\cite[Fig.~2, p.16]{IEEE149-1979}
\begin{equation}\label{eq:sphericalCoordinates}
\begin{aligned}
\vec{r}(r,\theta,\phi) &= r(\sin\theta\cos\phi\unitvec{x}+\sin\theta\sin\phi\unitvec{y}+\cos\theta\unitvec{z})\\
& = r\unitvec{r}(\theta,\phi),
\end{aligned}
\end{equation}
is used. Here, $\unitvec{x}$, $\unitvec{y}$ and $\unitvec{z}$ denote three orthogonal unit vectors of a right-handed $xyz$ Cartesian coordinate system.  Decompose the fields in transverse and radial parts, $\vec{E} = \vec{E}_t+E_r\unitvec{r}$, $\vec{H} = \vec{H}_t+H_r\unitvec{r}$, $\vec{S} = \vec{S}_t+S_r\unitvec{r}$. This leads to
\begin{equation}\label{eq:nearfieldPowerFlux}
\left\{
\begin{aligned}
\vec{S}_t &=\Re[\vec{E}_t\times H_r^*\unitvec{r} + E_r\unitvec{r}\times \vec{H}_t^* ]\\
S_r &=\Re[\unitvec{r}\cdot(\vec{E}_t\times\vec{H}_t^*)]
\end{aligned}	
\right.
\end{equation}
Note that only the tangential components contribute to the radial power flux density $S_r$. The presence of radial fields in the \nf\ region introduces a possible measurement error if the \acf{MA} and measurement setup have non-ideal characteristics, \eg\ non-planar wave fronts are measured or the \ac{MA} is mis-aligned. 
In the \ff\ $r\rightarrow\infty$, or for practical purposes $r\geq 2 D^2/\lambda$, the power density
\begin{equation}\label{eq:farfieldPowerFlux}
S_r \approx |\vec{E}_t|^2/Z_0,
\end{equation}
$Z_0$ being the free space wave impedance.
This relation holds also in most of the radiating \nf, cf. Section~\ref{sec:nearfield}. 

The \ac{TRP} is calculated from the power flux density data, by using the Poynting theorem~\cite{Cheng1989}, as
\begin{equation}\label{eq:trpFromPowerFlux}
\TRP = \iint\limits_{4\pi} S_r(r,\theta,\phi) r^2\dOmega.
\end{equation}
The integration is over the entire sphere\footnote{The intervals used to cover a full sphere do not matter for the results. Two common choices are $\theta\in[0,\pi]$, $\phi\in[-\pi,\pi]$ found in many text books, and the ``ball of yarn'' $\theta\in[0,2\pi]$, $\phi\in[0,\pi]$.}, solid angle $\Omega=4\pi$, and $r^2\dOmega=r^2\sin\theta\mathrm{d}\theta\mathrm{d}\phi$ is the infinitesimal surface area spanned by $\mathrm{d}\theta$ and $\mathrm{d}\phi $ on the sphere with radius $r$. In the \ff\ region the \ac{EIRP} is defined as

\begin{equation}\label{eq:eirpFromPowerDensity}
\mathrm{EIRP}(\theta,\phi) = \lim_{r\rightarrow \infty}S_r(r,\theta,\phi)4\pi r^2.
\end{equation}
This can be used to express \ac{TRP} as the angular average of the \ac{EIRP}
\begin{equation}\label{eq:trpFromEirp}
\TRP = \frac{1}{4\pi}\iint\limits_{4\pi}\mathrm{EIRP}(\theta,\phi)\dOmega
\end{equation}
This also follows from $\EIRP(\theta,\phi) = \TRP\, G_\mathrm{D}(\theta,\phi)$~\cite{IEEE145-2013}, where $G_\mathrm{D}$ is directivity and by definition 
\begin{equation}
\frac{1}{4\pi}\iint\limits_{4\pi} G_\mathrm{D}(\theta,\phi)\dOmega = 1.
\end{equation}
If power density is correctly measured in the \nf\ and~\eqref{eq:eirpFromPowerDensity} is used without taking the limit $r\rightarrow\infty$, the result will be proportional to power density flux, but not proper \ac{EIRP} values. However, if these results are used in~\eqref{eq:trpFromEirp} the resulting \ac{TRP} value will be correct.

The angular steps needed to correctly characterize the average value of a power density pattern is analyzed in Appendix~\ref{app:sampling}.  These steps are here denoted reference angular steps and are calculated as 

\begin{equation}\label{eq:refSteps}
\left\{
\begin{aligned}
\Delta\theta_\text{ref} &= \frac{\lambda/2}{R_\mathrm{sph}},\\
\Delta\phi_\text{ref} &= \frac{\lambda/2}{R_\mathrm{cyl}}
\end{aligned}
\right.
\end{equation}
Here, $R_\mathrm{sph}$ and $R_\mathrm{cyl}$  are the radii of the minimum sphere and $z$-axis centered cylinder, enclosing the \ac{EUT}, respectively. Note, from~\cite{Hald1998} it follows that the angular resolution in the radiating \nf\ is identical to the angular resolution in the \ff\ region. To quantify sampling on a sparse grid \iac{SF} is used. The \ac{SF} is defined as
\begin{equation}
\label{eq:sparsityFactorGeneral}
\mathrm{SF} = \max\left(\frac{\Delta\theta}{\Delta\theta_\mathrm{ref}},\frac{\Delta\phi}{\Delta\phi_\mathrm{ref}}\right).
\end{equation}
The maximum angular step is set as $15^\circ=\pi/12$ radians to be consistent with~\cite{CTIA2016}, which implies that 
5
\begin{equation*}
\SF\leq\SF_\text{max} = \pi R_\mathrm{sph}/6 \lambda.
\end{equation*}

\begin{figure}[]
\begin{center}
\includegraphics[width=\columnwidth]{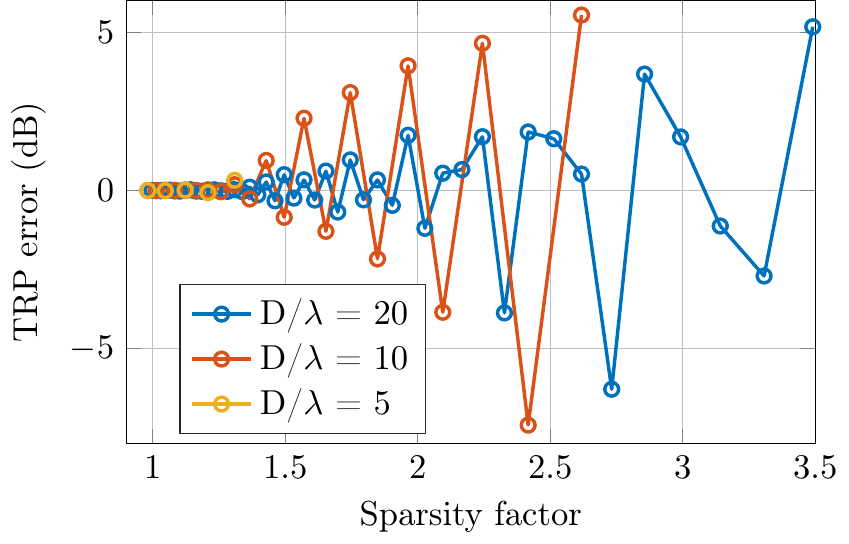}
\caption{Convergence of \ac{TRP} as the \ac{SF} is varied from 1 to 3.5, evaluated for different electrical sizes $D/\lambda$ and uniformly excited $N\times N$ element antenna arrays in the $yz$-plane. The number of rows $N=1+\ceil{\sqrt{2}D/\lambda}$ where $\ceil{\cdot}$ denotes rounding to nearest larger integer. This is the smallest $N$ for a uniform quadratic, array with element spacing $\leq \lambda/2$ that fits in a sphere of radius $D$, see Fig.~\ref{fig:UQA}.}
\label{fig:convergenceSFdOmega}
\end{center}
\end{figure}

The convergence of \ac{TRP} is shown in Fig.~\ref{fig:convergenceSFdOmega}. Moreover, $\Delta\theta_\mathrm{ref}= 15^\circ$ is equivalent to $D_\mathrm{sph}/\lambda = 12/\pi\approx 4$. Therefore, a distinction between small and large source dimensions is made with a split at $D=4\lambda$. Note that $\Delta\phi_\mathrm{ref}\geq\Delta\theta_\mathrm{ref}$ since $R_\mathrm{cyl}\leq R_\mathrm{sph}$.

\section{Sampling grid}\label{sec:sparsity}

According to~\eqref{eq:trpFromPowerFlux} and~\eqref{eq:trpFromEirp} 
\begin{equation}\label{eq:angularAverage}
\TRP = 4\pi \rmeas^2 \fsav{S_r} = \fsav{\EIRP},
\end{equation}
where $\fsav{\cdot}$ denotes full sphere angular average and $\rmeas$ is the radius of the measurement sphere. Thus, assessment of \ac{TRP} is about estimating full sphere averages. The accuracy of the estimate depends on: the angular grid type, the angular density of measurement samples, the antenna dimensions, and the degree of correlation between the sources\footnote{In the context of this paper the sources can be either the antenna elements or any current flowing on the \ac{EUT}.} of radiation. For highly correlated and separated sources the angular patterns can have narrow beams and for totally uncorrelated sources the beams of the pattern will be wider~\cite{Allen1960,Ruze1966}. The narrower the beams the smaller angular step is needed.

\begin{figure}[]
\begin{center}
\includegraphics[width=.49\columnwidth]{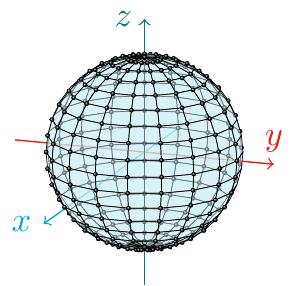}
\includegraphics[width=.49\columnwidth]{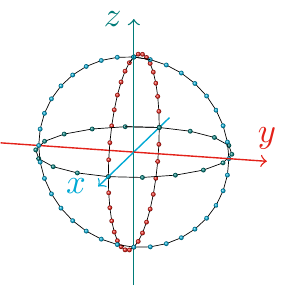}
\caption{Spherical grid (left) and three cuts grid (right). The mandatory two cuts are depicted in green and red.}
\label{}
\end{center}
\end{figure}
Two general classes of measurement grids are investigated: full-sphere, and multiple orthogonal cuts. Furthermore, for each grid type different angular resolutions are taken into account. The full-sphere grid is a rectilinear grid in the $\theta\phi$ plane. In grids of type multiple orthogonal cuts, two or three orthogonal cuts around the \ac{EUT} are used. The actual choice of grid type and the angular resolution is based on practical aspects such as turntable constraints and required measurement time.

The contribution from the grid to the \ac{TRP} is 
\begin{equation}\label{eq:gridAverage}
\TRP_\mathrm{grid} = 4\pi \rmeas^2 \gridav{S_r} = \gridav{\EIRP},
\end{equation}
where $\gridav{\cdot}$ denotes angular average over the used grid. For a full sphere the grid average is calculated as
\begin{equation}\label{eq:fullSphereAverage}
\gridav{u} = \frac{1}{4\pi}\sum_{m=1}^{N_\theta}\sum_{n=1}^{N_\phi} u_{mn}\Delta\Omega_m,
\end{equation}
where $\Delta\Omega_m$ is the solid angle per sample. Commonly this is calculated as $\Delta\Omega_m = \sin\theta_m \Delta \theta\Delta\phi$ but for sparse grids, by changing variable to $w=\cos\theta$ and using $\Delta\Omega = \Delta w_m\Delta\phi = \Delta(\cos\theta_m)\Delta\phi$ provides better numerical accuracy. For an orthogonal cuts grid
\begin{equation}\label{eq:orthogonalCutsAverage}
	\gridav{u} = \frac{1}{N}\sum_{p=1}^{N} \left\langle u^{(p)}\right\rangle_\text{cut},
\end{equation}
using $N=2$ or $3$ orthogonal cuts with samples $u^{(p)}$ in the $p$-th cut. To account for the error made when not measuring a dense full sphere an additional margin $\Delta \TRP$ is introduced as

\begin{equation}\label{eq:estimatedTrp}
\TRP_\mathrm{est} = \TRP_\mathrm{grid} + \Delta\TRP.
\end{equation}
For a dense full sphere grid, \ie\ $\SF=1$, $\Delta\TRP=0$. Note that for regulatory approval it is required that the \ac{TRP} is lower than certain limits. Therefore $\Delta\TRP$ must be chosen in a way that covers for cases that $\TRP_\mathrm{grid}$ is smaller than the \ac{TRP} in order to prevent false pass test results. On the other hand if $\TRP_\mathrm{grid}$ is an overestimate of the \ac{TRP}, no correction factor is needed.

\subsection{Electrically small sources}\label{subsec:smallAntennas}
Electrically small sources have dimensions of only a few wave lengths. Here the upper bound is set such that the corresponding reference angular step $\Delta\theta_\mathrm{ref}\geq 15^\circ$~\cite{CTIA2016} which is equivalent to $D\leq 12/\pi\approx 4\lambda$. Note that any \ac{EUT} can be regarded electrically small at sufficiently low frequencies.
 For this case an analysis based on \iac{SWE}~\cite{Hald1998} is used. A \ff\ amplitude pattern corresponding to sources within a sphere of diameter $\Dsph\leq 4\lambda$ correspond to a mode truncation at $l\leq kR=\frac{2\pi}{\lambda}2\lambda\approx 12=L$. The total number of modes is hence~\cite{Hald1998}
\begin{equation*}
J = 2 (L^2+2L) = 336.
\end{equation*}
Each statistical sample is generated by the following sequence
\begin{enumerate}
	\item Select number of modes $N$ such that $N\leq J$.
	\item Set modes to use by selecting $N$ integers randomly in the range $1,2,\ldots,J$. 
	\item Assign random weights $w_n = x_n+ \ju y_n$ where $x_n,y_n \in N(0,1)$, and normalize to unit \TRP\ by using~\eqref{eq:trpFromSwe}.
	\item Calculate $\EIRP(\theta,\phi)$ on the desired grids.
\end{enumerate}

Note that picking random modes implicitly selects random source rotations.

\begin{figure}
	\begin{center}
		\includegraphics[width=\columnwidth]{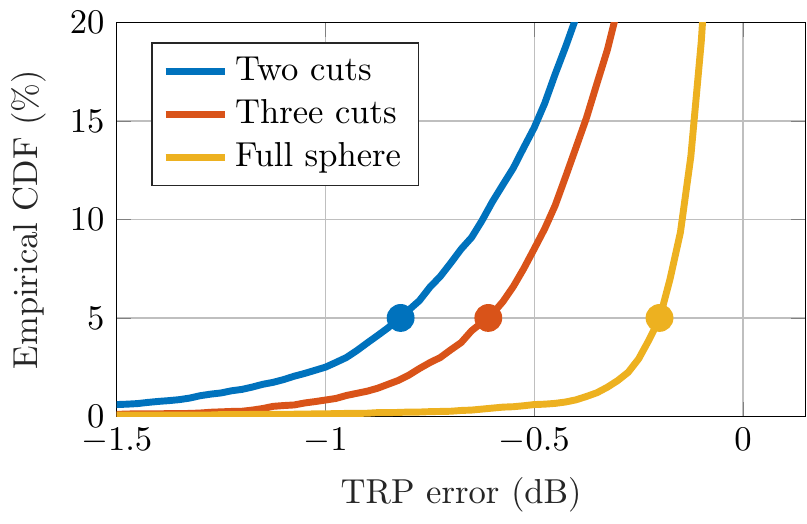}
		\caption{The TRP errors for small antennas ($D<4\lambda$) on different grids. The small error for full sphere method is caused by the numerical integration on small number of points on the sphere. The \ac{CDF} is even in the abscissa and the 50\% level is at 0 dB error.}
		\label{fig:smallAntennasTrp}
	\end{center}
\end{figure}

The statistics for 10,000 samples is depicted in Fig.~\ref{fig:smallAntennasTrp}. If a two cut grid is used $\Delta\TRP = 0.8$~dB with 95\% confidence level and $\Delta\TRP=0.2$~dB for full-sphere grid, indicated by dots in the figure. Note that the small error for full sphere grid is caused by the trapezoidal integration on small number of points on the sphere, whereas a more accurate integration scheme can reduce this error. The major conclusion is that a $15^\circ$ step full sphere grid accurately predicts the \TRP\ value for small antennas, cf~\cite{CTIA2016}.

%

\subsection{Electrically large sources}

Electrically large sources have large dimensions compared to a wavelength, and complement the case of electrically small sources by using the criterion $D\geq 4\lambda$. Note that any \ac{EUT} will be electrically large at sufficiently high frequencies.

\subsubsection{Low correlation}\label{sec:lowCorrelation}

For low correlation between sources a statistical approach is proposed. This is specifically addressing spurious emissions of less known nature, and one idea is to take height for an arbitrary device orientation. As a direct consequence there will be no need for aligning the \ac{EUT} to the emission pattern. The following notation is used: $N(0,1)$ denotes a normal distribution of zero mean and unit standard deviation, and $U(a,b)$ denotes a uniform distribution on the interval $[a,b]$.

Emissions with low correlation can have patterns with peaks in arbitrary directions. Hence, aligning the emission peaks to the measurement grid can be difficult. On the other hand, if the correlation is low the angular resolution will be low~\cite{Allen1960,Ruze1966} and the alignment will play a minor role. Therefore, to avoid the need to align the \ac{EUT} to the power flux density pattern, and to investigate the effect of sparse sampling and choice of grid type for uncorrelated emissions, a statistical approach is proposed. \ac{TRP} is calculated for a large number of array antennas of a given electrical size $D/\lambda$. The statistical samples include random rotations, see App.~\ref{app:rotations}, and hence the end result will be valid for any rotation of the \ac{EUT}. To comply with Sec.~\ref{subsec:smallAntennas} the maximum angular step is $15^\circ$ and only dimensions $D\geq 4\lambda$ are considered.

For electrically large sources the angular resolution is dominated by the array factor, \ie, only the positions of the sources of radiation will contribute. For this reason a point source model is used and the \ac{EIRP} pattern is calculated as
\begin{equation}\label{eq:eirpPointSources}
\EIRP(\theta,\phi) = \left|\sum\limits_{n=1}^N  \eu^{\ju k\unitvec{r}(\theta,\phi)\cdot\vec{d}_n}w_n\right|^2.
\end{equation}
Here, $\vec{d}_n$ are the positions of the $N$ radiation sources,  $w_n$ are the complex amplitudes of each source, and $\unitvec{r}(\theta,\phi)$ is the radial unit vector in the \ac{SSCS} of Eq.~\eqref{eq:sphericalCoordinates}.  For the sake of simplicity, \EIRP\ patterns and an infinite test distance is used.

\begin{figure}
	\begin{center}
		\includegraphics[width=5cm]{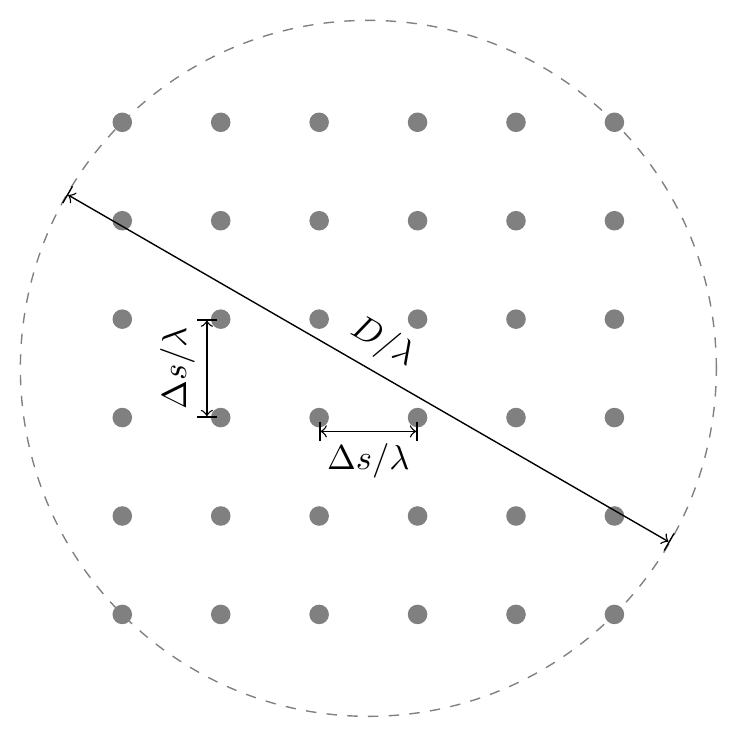}
		\caption{The array model. A uniform quadratic array inscribed in a sphere of diameter $D$. To ensure that $\Delta s\geq \lambda/2$ the maximum number of rows is $1+\sqrt {2}(D/\lambda)$. }
		\label{fig:UQA}
	\end{center}
\end{figure}

The statistical analysis is based on randomly rotated quadratic arrays of a certain electric size $D/\lambda$, see Fig.~\ref{fig:UQA}. Hence, $D=\Dsph=\Dcyl$ and $\Delta\theta_\text{ref}=\Delta\phi_\text{ref}=\lambda/D$. Source excitations with correlation $\rho$ are calculated using
\begin{equation}\label{eq:correlationParametrization}
w_n(\rho) = \sqrt{\rho} + \frac{(x_n+\ju y_n)}{\sqrt{2}}\sqrt{1-\rho},
\end{equation}
where $x_n,y_n\in N(0,1)$. This implies that $\overline{w_m^*w_n}=1$ if $m=n$ and $\rho$ if $m\neq n$.

The samples of the statistical analysis are generated by the following steps:
\begin{enumerate}
\item Make a uniform quadratic array of source points with $N_\text{row}\times N_\text{row}$ elements and diameter $D$, see Fig.~\ref{fig:UQA}, where $N_\text{row}$ is a random integer in the range $2$ to $10$ and the source separation $\Delta s\geq\lambda/2$. Rotate the source positions by using a random rotation matrix $\mat{R}_\text{rot}(\alpha,\beta,\gamma)$, see Appendix~\ref{app:rotations}.
\item Calculate source weights using \eqref{eq:correlationParametrization} where $\rho\in U(0,\rho_\text{max})$. Normalize the weights such that $\TRP=1$ on a \emph{dense grid}.
\item Calculate $\TRP_\text{grid}$ for the grid and angular step, $\Delta\theta<15^\circ$, of interest by using \eqref{eq:gridAverage} and \eqref{eq:eirpPointSources}.
\end{enumerate}

\begin{figure}
	\begin{center}
		\includegraphics[width=\columnwidth]{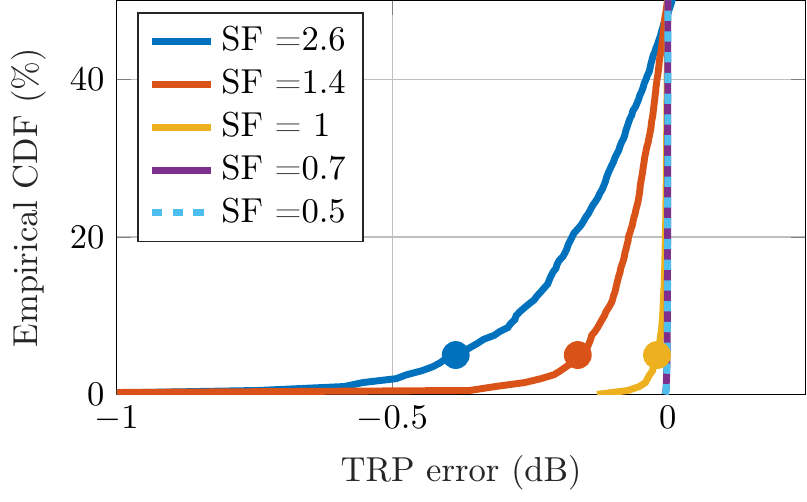}
		\caption{Empirical CDFs for $D=10\lambda$,  $\rho_\mathrm{max}=0.2$, and full-sphere grid for different \acp{SF}, based on $10^4$ samples. For $\text{SF}\leq 1$ the TRP error is negligible at all levels, and for $\SF=1$ the error is negligible at the 5\% percentile. The reference step $\Delta\theta_\mathrm{ref}\approx 2.9^\circ$. The \ac{CDF} is even in the abscissa and the 50\% level is at 0 dB error.}
		\label{fig:convergence10Lambda}
	\end{center}
\end{figure}

Note, several grids and angular steps can be analyzed in parallel in step three.  Furthermore, the approach described here is not limited to the grids investigated here. A number of $10^4$ samples are used to ensure statistical convergence. If a negative \ac{TRP} error is found at the 5\% percentile then its absolute value is used as $\Delta\TRP$, otherwise $\Delta\TRP=0$. This will ensure a 95\% confidence of the estimated TRP~\eqref{eq:estimatedTrp}. Dots indicate the 5\% percentile values in Fig.~\ref{fig:convergence10Lambda}.

Empirical \acp{CDF} for $D=10\lambda$, $\rho_\mathrm{max}=0.2$, and a full sphere grid, are depicted in Fig.~\ref{fig:convergence10Lambda}, showing that the \ac{TRP} error is negligible for $\SF\leq 1$ with 95\% confidence. Note that this result depends slightly on the numerical method~\eqref{eq:fullSphereAverage} used to calculate the full sphere angular average. 

\begin{figure}
	\begin{center}
		\includegraphics[width=0.99\columnwidth]{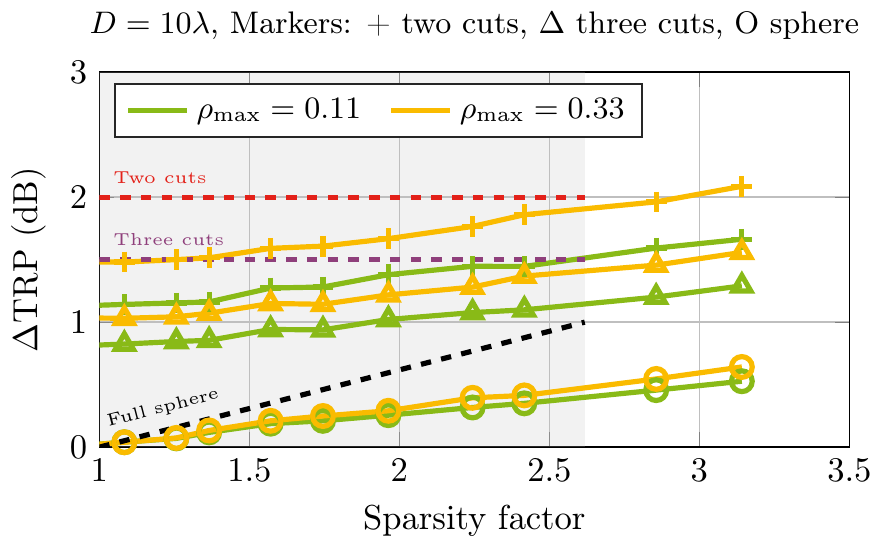}\\(a)\\
		\includegraphics[width=0.99\columnwidth]{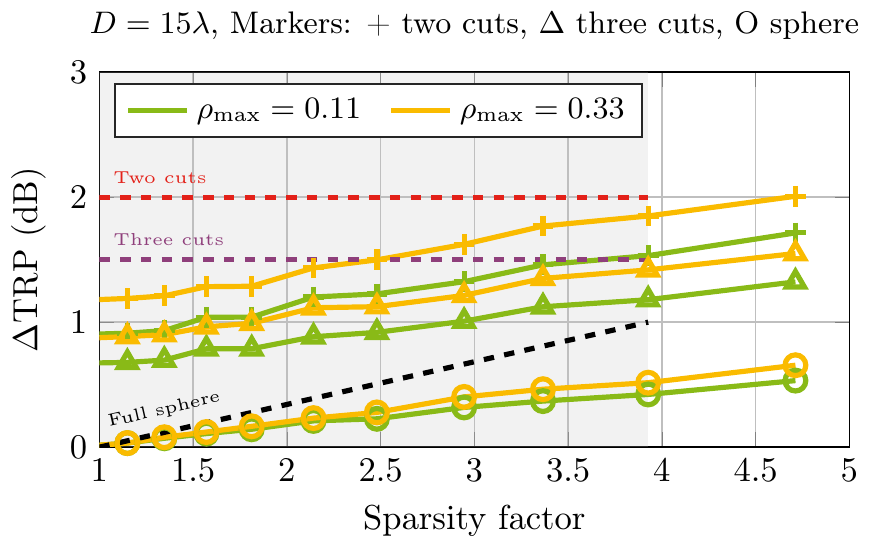}\\(b)\\
		\includegraphics[width=0.99\columnwidth]{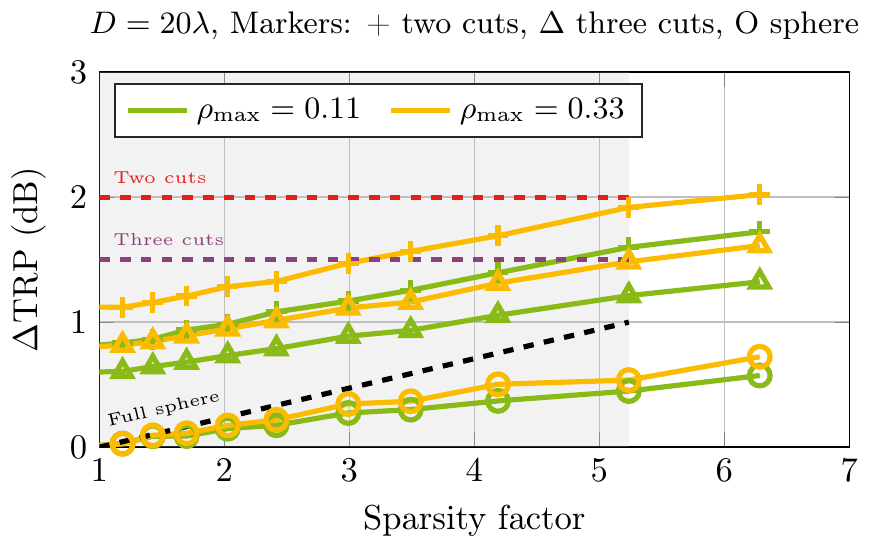}\\(c)
		\caption{$\Delta$TRP vs SF for different electrical sizes ($D/\lambda$) and correlations ($\rho_\mathrm{max}$) using 10,000 samples. The gray area highlights the angular sampling $\Delta\theta\leq 15^\circ$.}
		\label{fig:deltaTRP}
	\end{center}
\end{figure}

Values of $\Delta\TRP$ for different electrical sizes $D/\lambda$ and max correlations $\rho_\text{max}$ are shown in Figs~\ref{fig:deltaTRP} a-c. Three major trends can be observed. Firstly, the two cuts grid has the highest $\Delta\TRP$, followed by three cuts and full sphere. This is expected as the spherical coverage gets more uniform by following this sequence. Secondly, $\Delta\TRP$ increases with \ac{SF}, since fewer angular points are used to calculate $\TRP_\text{grid}$. Thirdly, $\Delta\TRP$ increases with the correlation, which is due to the fact that the array factor dominates over the element pattern~\cite{Allen1960,Ruze1966} when the correlation increases. This results in a need for denser angular grids. A closer look at these figures shows a similar trend for different values of $D$, although with different scaling, \ie, the horizontal axis. This can be explained by the fact that the correlation between sources are low and the element pattern has a dominant role.

Based on a large investigation with many sizes and correlation levels, a subset is presented in Figs~\ref{fig:deltaTRP} a-c, an upper bound for $\Delta\TRP$ is proposed. The upper bounds are shown in the same figures with dashed lines, and are summarized in Table~\ref{tab:deltaTrpUncorrelated}. The small slope of the curves suggests flat upper bounds, but for the full sphere, an inclining upper bound is proposed to allow for $\Delta\TRP=0$ at $\SF=1$. Hence, for two and three cuts the proposed values for $\Delta\TRP$ are $+2$dB and $+1.5$ dB, respectively, whereas for the full sphere $\Delta\TRP = (\SF-1)/(\SF_\text{max}-1)$ dB. Note that $\SF_\text{max}$ corresponds to $\Delta\theta=\Delta\phi=15^\circ$.

\begin{table}
\begin{center}
\caption{Proposed values of $\Delta\TRP$ for uncorrelated emissions.}
\label{tab:deltaTrpUncorrelated}
\begin{tabular}{cccc}
Grid type&Two cuts&Three cuts& Full sphere\\\toprule
$\Delta \TRP$ (dB) &2 &1.5& $(\SF-1)/(\SF_\text{max}-1)$\\\bottomrule
\end{tabular}
\end{center}
\end{table}

\subsubsection{Correlated sources}
For correlated emissions the straightforward solution is to measure power flux density on a \emph{dense sphere}, $\SF=1$, which may result in very lengthy measurements. Two ways to reduce the measurement time have been identified. Either, the pattern lobes are narrow but the symmetries can be exploited, or the beams can be effectively widened by using a beam sweeping test signal. If the symmetries of the pattern is known, and the angular grid is aligned to the cardinal cuts, then an overestimate of the \ac{TRP} is found. This is most probably the case for emissions at frequencies close to the operating band. Furthermore, if the symmetries of the antenna is known the power flux density pattern is well characterized by measuring two cardinal cuts, and then use \iac{PM} technique to retrieve values outside the cardinal cuts in order to get full sphere data. This is further described in the next paragraph, and the the beam sweeping test signal is further investigated in Sec.~\ref{sec:config}. 

The proposed \ac{PM} is based on the possibility to calculate the array factor of a rectangular array as a product of two terms, corresponding to two orthogonal cuts. Assuming an array positioned in the $yz$-plane, this must be done in two separate forward and backward hemispheres. Therefore, the data is separated in two hemispheres and the estimated radiated powers are added together. The two hemispheres are defined as\footnote{Note that the $\sin\theta$ factor is needed since a full turn $\theta\in[0,2\pi]$ is used for the vertical cut.}
\begin{equation*}
	\sin\theta\cos\phi
	\begin{cases}
\geq0&\text{forward (fwd),}\\
\leq 0&\text{backward (bwd).}
\end{cases}
\end{equation*}
To exploit the rectangular array geometry, the \ac{PM} uses the coordinates
\begin{equation}
\left\{
	\begin{aligned}
		u &= y/r = \sin\theta\sin\phi,\\
		v &= z/r = \cos\theta.
	\end{aligned}
\right.
\end{equation}
The needed data are the horizontal and vertical power flux densities $S_r^H(u)=S_r(u,0)$ and $S_r^V(v)=S_r(0,v)$, respectively. The power flux density in a point $(u,v)$ is calculated as
\begin{equation}
	S_r(u,v) = \frac{S_r^H(u)S_r^V(v)}{S_r(0,0)}
\end{equation}
where $S_r(0,0)=S_r^H(0)=S_r^V(0)$ is the power density at the crossover point. Note that power density at the crossover point is measured in both cuts and these values must be equal with a reasonable accuracy. The \ac{TRP} is calculated as
\begin{equation}
\begin{aligned}
	\TRP = &\iint_{u^2+v^2\leq 1} S^\text{fwd}_r(u,v)  \rmeas^2 \dOmega\\
	& + \iint_{u^2+v^2\leq 1}  S^\text{bwd}_r(u,v)  \rmeas^2 \dOmega.
\end{aligned}
\label{eq:uvIntegration}
\end{equation}
Where $\dOmega(u,v)=\mathrm{d}u\mathrm{d}v/\sqrt{1-(u^2+v^2)}$. Note that $\dOmega$ is singular at $u^2+v^2=1$ which must be taken care of in the integration, \eg\ as shown in Appendix~\ref{app:integration}.

\begin{figure}
	\begin{center}
		\includegraphics[width=0.75\columnwidth]{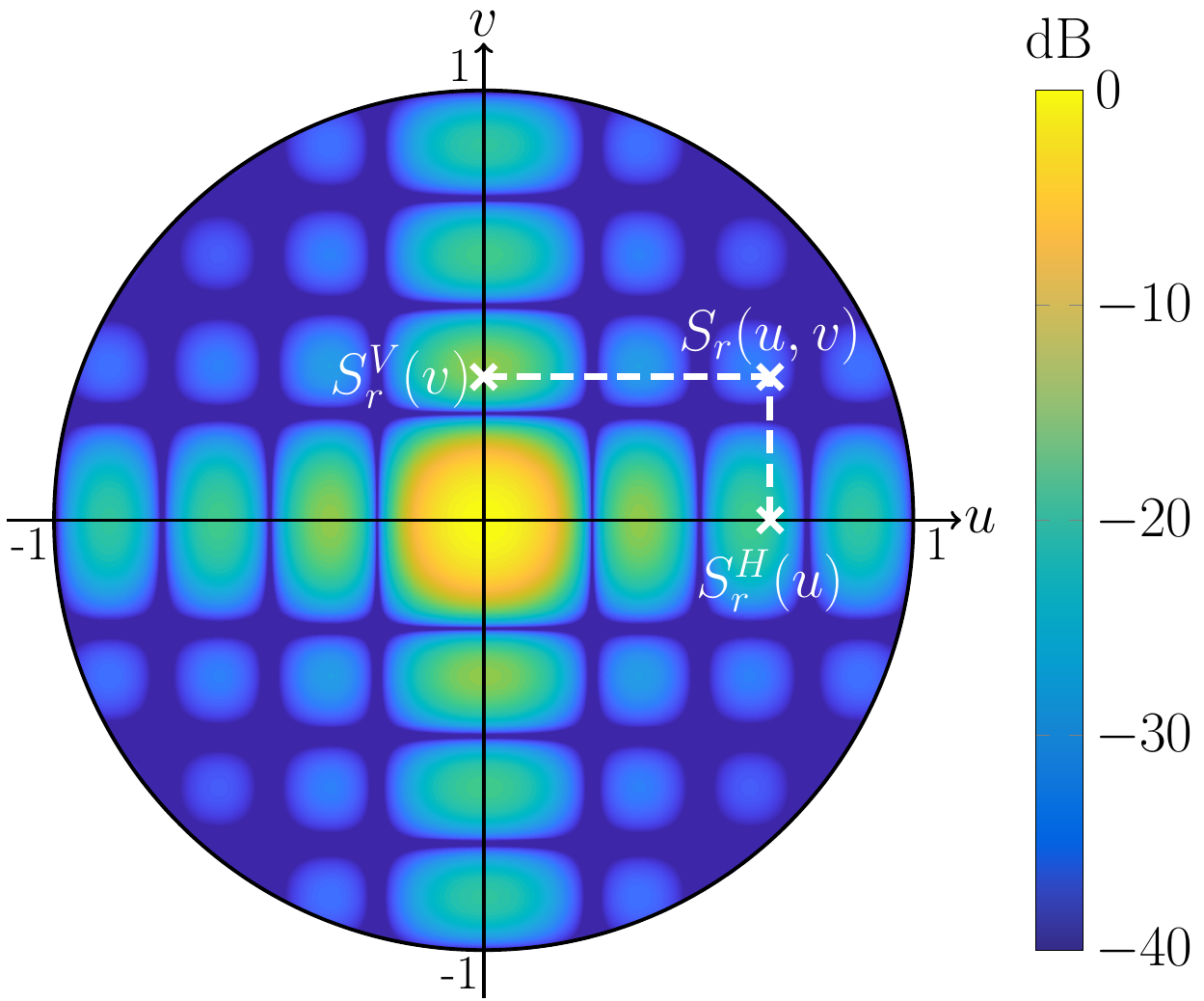}
		\caption{Radial power flux density pattern, in relative gain scale, of an $8\times 8$ array of $z$-oriented half-wave dipoles located in the plane $x=0$. The pattern is plotted in the $uv$ coordinate system. Pattern multiplication uses the on-axis values to calculate off-axis values as indicated with the white crosses and dashed lines.}
		\label{fig:pattMult}
	\end{center}
\end{figure}

To illustrate the \ac{PM} method, an $8\times 8$ array of $z$-oriented half-wave dipoles in the $yz$-plane is used. The radiation pattern of this array in the $uv$-plane is shown in Fig.~\ref{fig:pattMult} where the advantage of transformation to $uv$-plane is evident as well. If~\eqref{eq:orthogonalCutsAverage} is used an overestimation of almost 9~dB will result, whereas applying \ac{PM} will reduce the error to virtually 0~dB.

\section{Worst case configuration}\label{sec:config}
For the \ac{TRP} assessment of a device with many possible antenna configurations finding the worst case can be a practical difficulty. At different frequencies different configurations can lead to radiation in unpredictable directions with different levels of power. The only way to determine the worst configuration for each frequency would be to perform the measurement independently for each configuration in an exhaustive search manner. This can be a very time-consuming procedure. Even if the worst case configuration is known, it can be argued that this static state is not a representative mode for an \ac{AAS} with dynamic beam forming and beam tracking capabilities.

\begin{figure}
	\begin{center}
		\includegraphics[width=1.0\columnwidth]{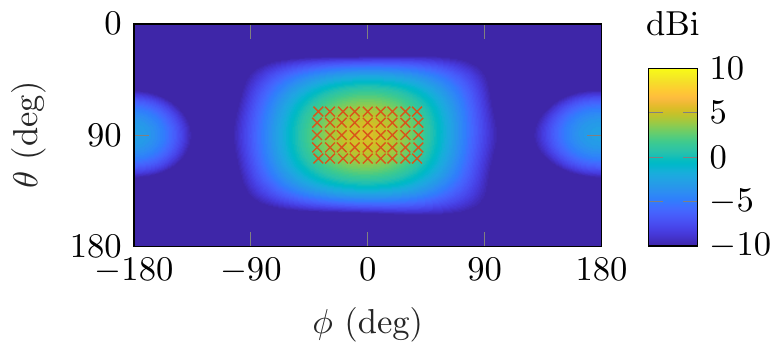}\\
		\includegraphics[width=1.0\columnwidth]{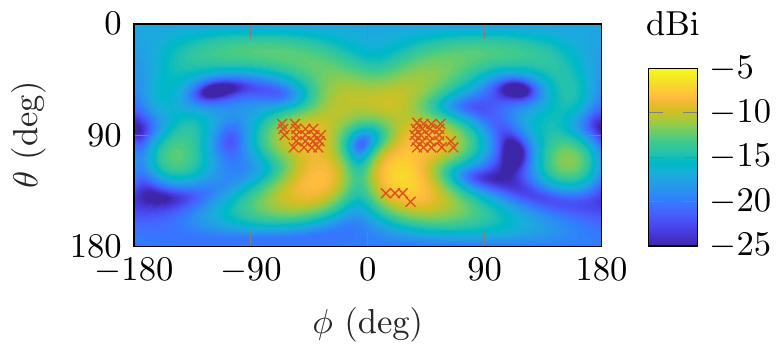}\\
		\includegraphics[width=1.0\columnwidth]{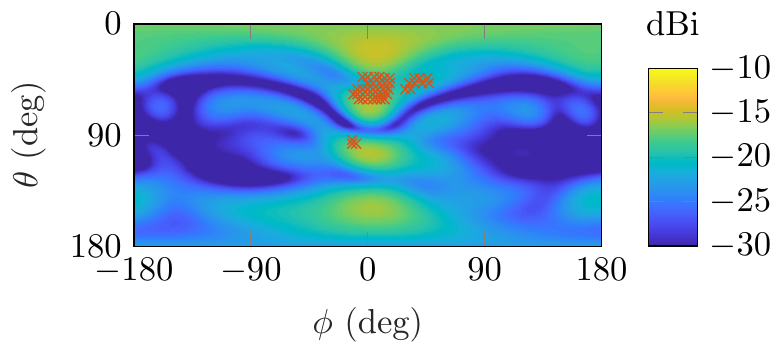}
		\caption{The embedded element pattern and the direction of the peaks of the array at different frequencies (top) fundamental frequency, (center) second harmonic, and (bottom) third harmonic.}
		\label{fig:beamDirections}
	\end{center}
\end{figure}

\begin{figure}
	\begin{center}
		\includegraphics[width=0.9\columnwidth]{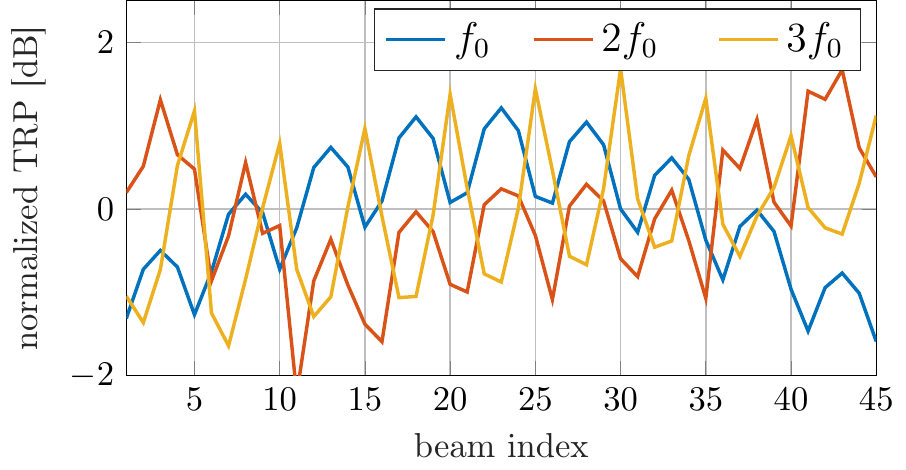}
		\caption{TRP corresponding to each beam at harmonic frequencies. The values are normalized such the the mean value is to 0 dB for every frequency.}
		\label{fig:beamTRPs}
	\end{center}
\end{figure}

As an illustration of unpredictable beam directions an example is provided. Assume an antenna array with 45 predefined beam positions on a $9\times 5$ grid. The beams span $80^\circ$ in the azimuth plane, $40^\circ$ in elevation and are uniformly spaced with $10^\circ$ between adjacent beams. The embedded radiation patterns of the antenna elements are obtained from \ac{FEM} simulations in the presence of nearest neighbor elements. Beam positions are studied at the second and third harmonics. In the absence of a complete model for out-of-band phase noise the same excitation is used at all frequencies. In Fig.~\ref{fig:beamDirections} the location of the beam peaks are depicted using the embedded element pattern, the envelope of the steered beams, as background. In Fig.~\ref{fig:beamTRPs} the \ac{TRP} of the beams are shown. The \ac{TRP} values are normalized such that the mean value is at 0 dB for every frequency. It is observed that not only the peak directions, but also the index of the worst case beam is frequency dependent.

Instead of searching for and using a worst case configuration, a sweep through all the predefined beams can be used. During the measurement process the beam sweeping average of the radiated power flux density is measured at each measurement point. This averaging matches well with the fact that \ac{TRP} is a spatial average of the measured values at different directions. Note, if instead the maximum value would be recorded for each point during the beam-sweeping, the resulting \ac{TRP} estimate will be an overestimate by a wide margin. The resulting \ac{TRP} value can afterwards be adjusted by a peak-to-average value which appears to be fairly similar for all frequencies as observed in Fig.~\ref{fig:beamTRPs}.

\begin{figure}
    \begin{center}
        \includegraphics[width=0.95\columnwidth]{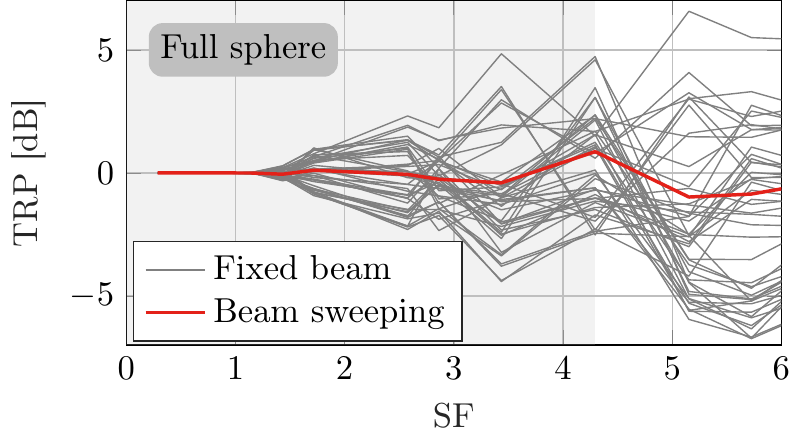}\\
        \includegraphics[width=0.95\columnwidth]{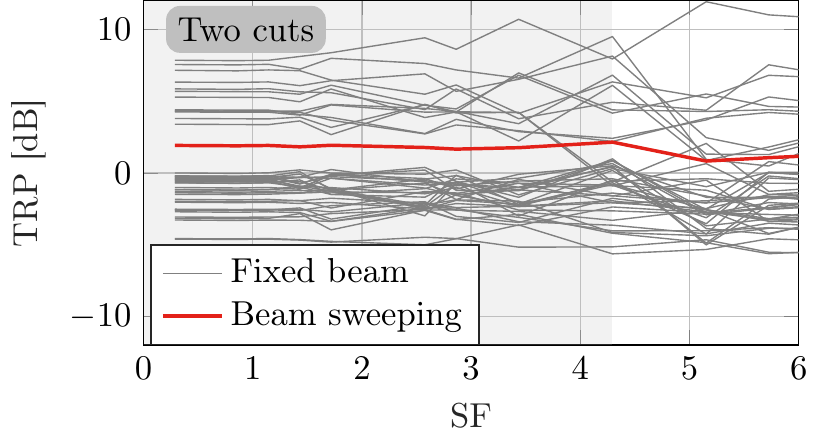}
        \caption{Sparse sampling, beam configurations, and beam sweeping configuration for different grids. The gray area highlights $\Delta\theta\leq15^\circ$.}
        \label{fig:convergenceBeamSweeping}
    \end{center}
\end{figure}

The effect of using a beam sweeping test signal on the convergence of sparse sampling is shown in Fig.~\ref{fig:convergenceBeamSweeping} as the error in \ac{TRP} vs. \SF, for full sphere and two cuts grids at the third harmonic. For both grids it is clear that the beam scanning test signal will result in smaller errors for larger \SF\ which means that fewer number of samples can be used. This reduction in the number of samples, combined with the fact that only one configuration must be tested, will significantly reduce the total testing time. Regarding the two cut grid, it is noteworthy that the individual beams can lead to errors as large as 10~dB, while the beam scanning average leads to an overestimation error of almost 2~dB regardless of \SF.

\section{\NF\ radial power flux density}\label{sec:nearfield}

\NF\ effects on \ac{TRP} is divided in two parts. First, it is investigated in which spatial domain the \ff\ expression for radial power flux density \eqref{eq:farfieldPowerFlux} can be used. Secondly, the measurement errors associated with measurement of this quantity in the \nf\ region are discussed.

\subsection{Validity of \ff\ approximation}\label{sec:deviationFarfieldApprox}

To evaluate the effect of using the \ff\ approximation of the radial power flux density \eqref{eq:farfieldPowerFlux}, a back propagation technique based on \iac{SWE}~\cite{Hald1998,Blanch2002,Friden2003} is used. This method enables calculation of $\vec{E}$ and $\vec{H}$ fields including radial components, and using \ff\ data as input, see App.~\ref{app:swe} and~\cite{Hald1998}. The fields can be calculated in any point excluding the zone $\rmeas\leq \Rsph+\Delta R$ where $\Delta R$ is on the order of $1\lambda$. Hence, the \ac{SWE} provides a tool to compare the true near field expression~\eqref{eq:nearfieldPowerFlux} and the far field approximation~\eqref{eq:farfieldPowerFlux} of radial power flux density.

\begin{figure}
\begin{center}
\includegraphics[width=\columnwidth]{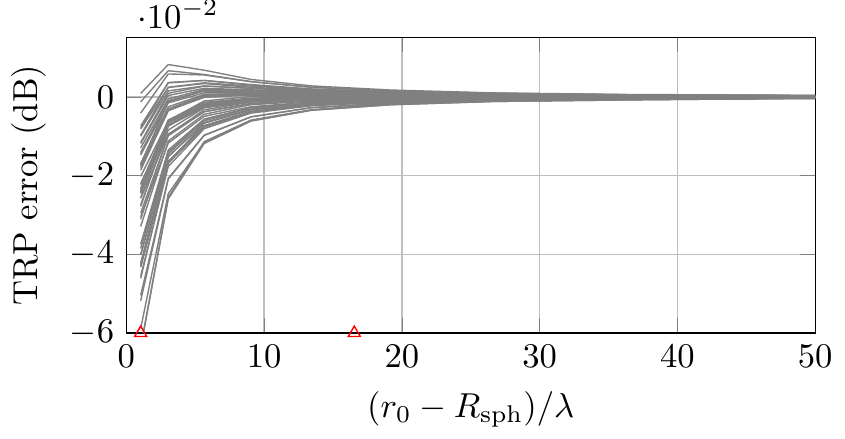}
\caption{\ac{TRP} error for different configurations of a base station antenna using the approximate far field expression \eqref{eq:farfieldPowerFlux} and comparing with the correct near field radial power flux density \eqref{eq:nearfieldPowerFlux}. Red triangles mark the distances 75~cm and 2.5~m, see Fig.~\ref{fig:bsaDistances}.}
\label{fig:macro2TrpError}
\end{center}
\end{figure}

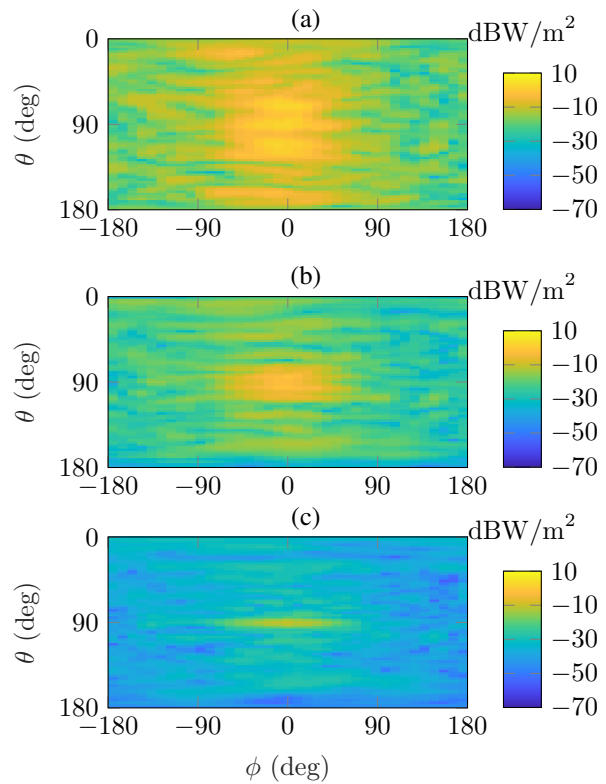
\begin{figure}
	\begin{center}
	\begin{tikzpicture}
		\node (a) at (0,7.0) {\includegraphics[width=8cm,clip=true,trim=0 21 0 0]{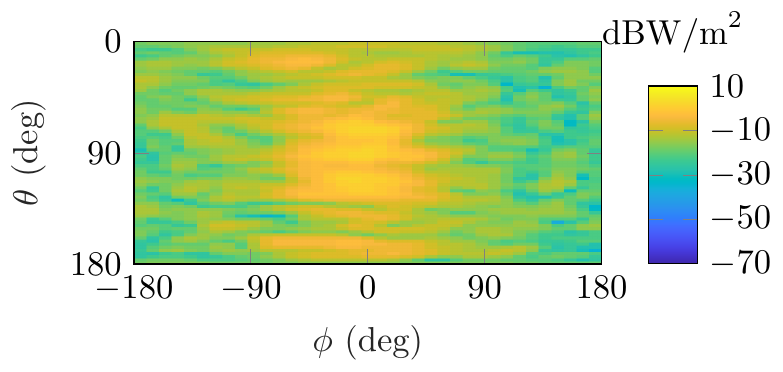}};
		\node (b) at (0,3.5) {\includegraphics[width=8cm,clip=true,trim=0 17 0 0]{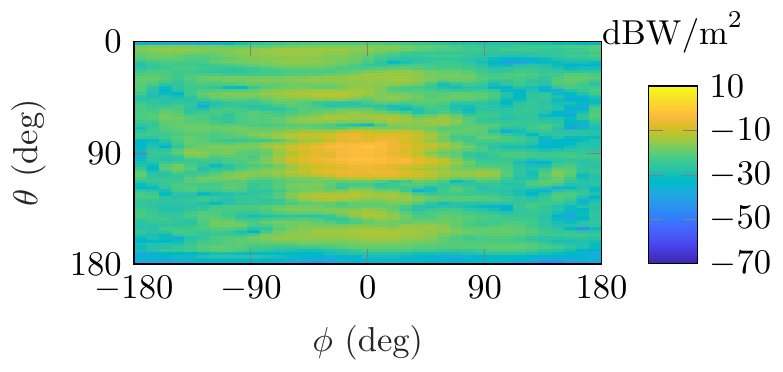}};
		\node (c) at (0,0) {\includegraphics[width=8cm]{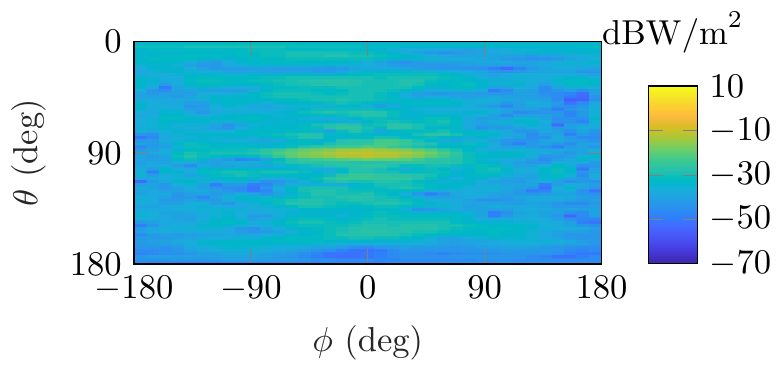}};
		\node at ([yshift=-3mm]a.north) {(a)};
		\node at ([yshift=-3mm]b.north) {(b)};
		\node at ([yshift=-3mm]c.north) {(c)};
      \end{tikzpicture}
		\caption{Pattern of radial power flux density for a base-station antenna with spherical radius of 63 cm, calculated by \eqref{eq:nearfieldPowerFlux} at different distances $\rmeas$ (top) 75 cm, (center) 2.5 m, and (bottom) 30 m. While the pattern and the peak level changes, the \ac{TRP} (1W) is the same at all distances.}
		\label{fig:bsaDistances}
	\end{center}
\end{figure}

As a first example, a four column macro base station antenna with $\Rsph=0.65$~m at 2655~MHz is considered. 56 curves corresponding to four antenna ports and 13 different tilt settings are depicted in Fig.~\ref{fig:macro2TrpError}. The red triangles at $\Rsph+\lambda\approx 0.75\text{m}$ and $\Rsph+17\lambda\approx 2.5\text{m}$ indicate the distances used in Figs~\ref{fig:bsaDistances} a-b showing the corresponding full sphere patterns. In this case, the \ff\ expression \eqref{eq:farfieldPowerFlux} can be used with an error below 0.065~dB all the way into 0.75~m.  While a distinct sector beam appears in the \ff\ pattern, the \nf\ patterns have a different shape. While the radial power flux density is proportional to the $\EIRP$ only in the \ff\ pattern, at all distances $\rmeas\geq 0.75$~m the \ff\ power flux density approximation~\eqref{eq:farfieldPowerFlux} is close to the true expression \eqref{eq:nearfieldPowerFlux}. Note that by energy conservation the \ac{TRP} is ideally independent of distance.  

\begin{figure}[]
\begin{center}
\includegraphics[width=\columnwidth]{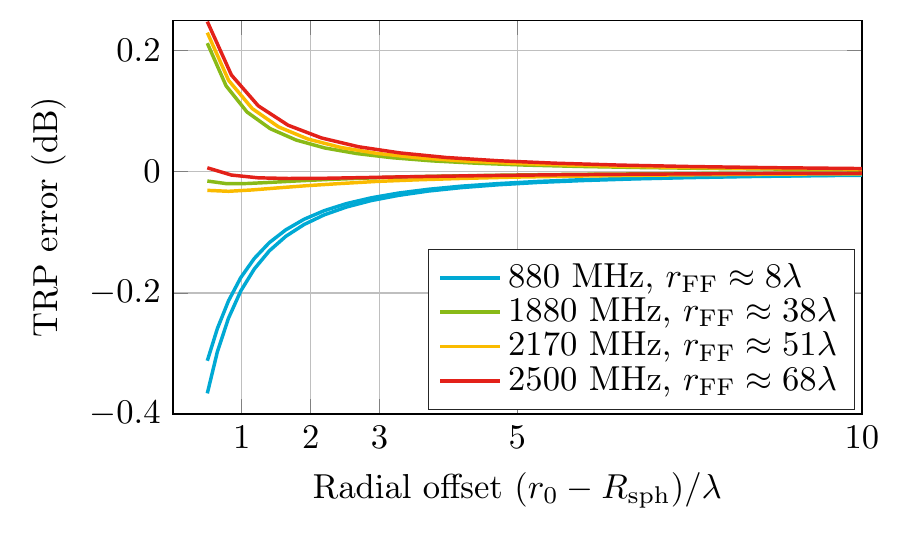}
\caption{Relative error in \ac{TRP} for two laptop antennas using the approximate \ff\ expression \eqref{eq:farfieldPowerFlux} and comparing with the correct nearfield radial power flux density \eqref{eq:nearfieldPowerFlux}.}
\label{fig:notebookTrpError}\end{center}
\end{figure}

A second example depicts the \ac{TRP} error for two antennas on a laptop as seen in Fig.~\ref{fig:notebookTrpError}. Since a comparison between the two antennas is not intended, the two are plotted with the same color at different frequencies. In this case a TRP error below 5\% is achieved by using the \ff\ approximation at $\rmeas=R+\lambda$, and below 1\% at $\rmeas = \Rsph+3\lambda$. The far field distances $r_\text{FF}$ are based on the mechanical dimensions of the laptop computer. 

\begin{figure}[]
\begin{center}
\includegraphics[width=\columnwidth]{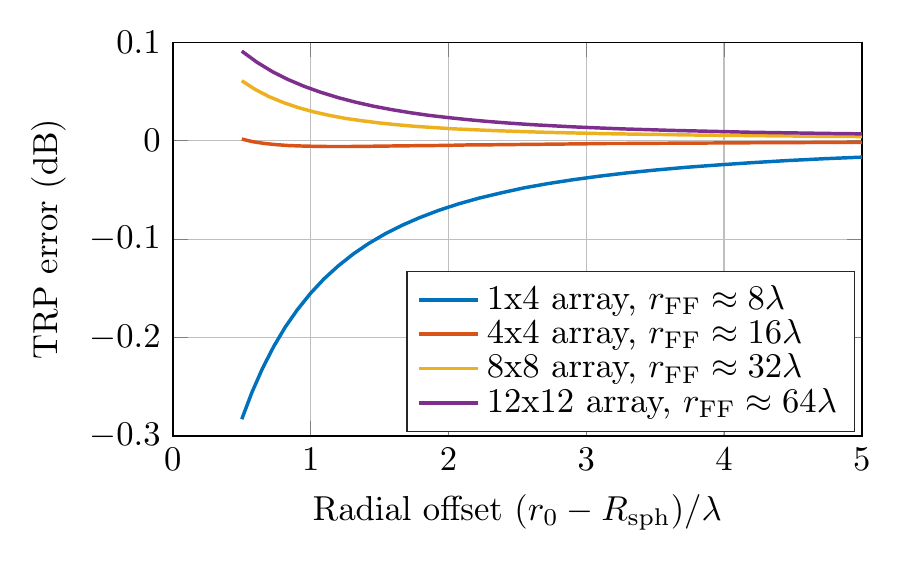}
\caption{Relative error in TRP calculation when using the \ff\ radial power flux density approximation~\eqref{eq:farfieldPowerFlux} for arrays of Hertz dipoles.}
\label{fig:trpErrorDipoles}
\end{center}
\end{figure}

In a last example, arrays of vertical electric Hertz dipoles are used. The relative error in \ac{TRP} is depicted in Fig.~\ref{fig:trpErrorDipoles}, and the relative error is below 0.05 dB at distances $\rmeas\geq \Rsph+3\lambda$. As in the previous examples, the far field distance $2D^2/\lambda$ seems to be an irrelevant parameter when evaluating the error caused by using the \ff\ radial power flux density approximation~\eqref{eq:farfieldPowerFlux} in the \nf.

Table~\ref{tab:errors} shows a comparison between the errors caused by back-propagation and those of the power flux approximation, for all the aforementioned examples. For each case the largest error is presented. The back propagation error is at least one order of magnitude smaller than the power flux approximation error, which verifies that the errors presented in Figs~\ref{fig:macro2TrpError},\ref{fig:notebookTrpError}, and \ref{fig:trpErrorDipoles} are not numerical errors due to back propagation.

\begin{table}
    \begin{center}
    \caption{Errors in estimation of TRP in near-field}
    \label{tab:errors}
    \begin{tabular}{p{0.3\columnwidth} p{0.3\columnwidth} p{0.25\columnwidth}}
                             & \begin{sloppypar}Back propagation (dB)\end{sloppypar}&\begin{sloppypar}Power~flux density approximation (dB)\end{sloppypar}\\ \toprule
        base station antenna & 5.5e-3                      & 6.5e-2 \\
        notebook antenna     & 2.6e-3                      & 2.5e-1 \\
        1x4 array            & 5.5e-6                     & 2.8e-1 \\
        4x4 array            & 5.3e-6                     & 1.8e-3 \\
        8x8 array            & 6.6e-5                     & 6.1e-2 \\
        12x12 array          & 1.1e-4                     & 9.1e-2 \\ \bottomrule
    \end{tabular}
    \end{center}
\end{table}

\subsection{Measurement antenna considerations}
\begin{figure}[]
\begin{center}
\includegraphics[width=\columnwidth]{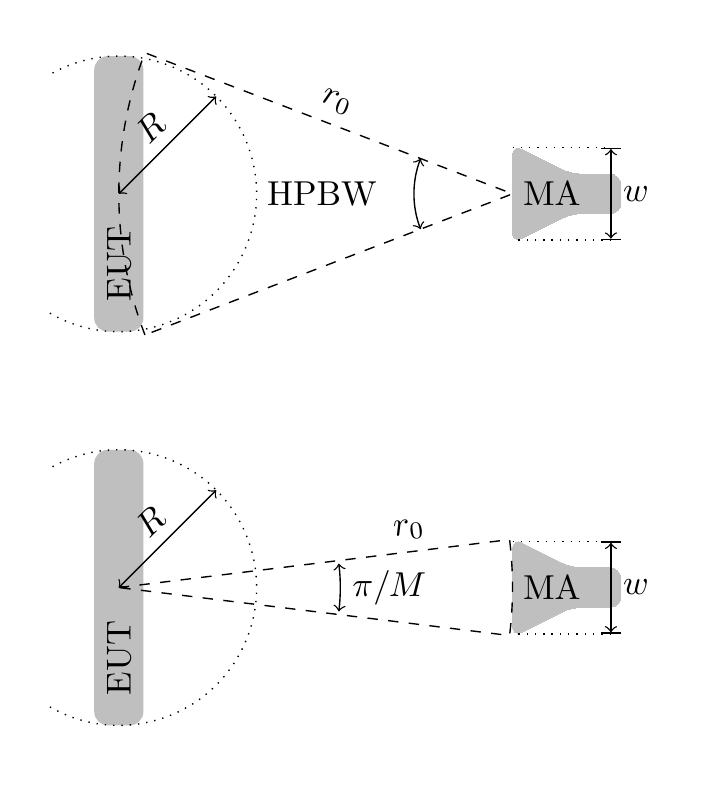}
\caption{\acf{MA} width criteria using \protect\ac{HPBW} (upper figure) and the angular resolution $\Delta\phi=\pi/M$ (lower figure).}
\label{fig:probeWidth}
\end{center}
\end{figure}

The results of Sec.~\ref{sec:deviationFarfieldApprox} imply that \ac{TRP} can be assessed from $|\vec{E}_t|$ if only the test distance $\rmeas$ exceeds $\Rsph$ by a few $\lambda$. To measure field data accurately, $\rmeas$ and the \ac{MA} must be handled appropriately~\cite{Yaghjian1986,IEEE149-1979}. To suppress influence from radial field components, the \ac{MA} should be carefully aligned and the \ac{EUT} should be placed in the \ff\ region of the \ac{MA}. Furthermore, the \ac{HPBW} of the \ac{MA} shall cover the \ac{EUT}, and to avoid influence from chamber scattering, excessive coverage shall be avoided~\cite{IEEE149-1979}. Hence, using the relation $\hpbw = \beta\lambda/w$~\cite{Balanis1997} and the coverage criterion $2R\leq \rmeas\:\hpbw$, cf. upper part of Fig.~\ref{fig:probeWidth}, implies that
\begin{equation}\label{eq:dminHpbw}
\rmeas \geq \frac{R}{\beta}\frac{w}{\lambda/2}
\end{equation}
Here $w$ is the width of the \ac{MA} and $\beta\approx 1.2$ for an open ended waveguide or \ac{SGH}.
An alternative approach is based on the angular resolution of the \ac{EUT} field. The received voltage at the \ac{MA} port is modeled as the reaction integral~\cite{Kildal2015}
\begin{equation}\label{eq:reactionIntegral}
V\propto \iint\limits_A \vec{E}\cdot\vec{J}_a\mathrm{d}A.
\end{equation}
Here $\vec{J}_a=\unitvec{n}\times\vec{H}_a$ is the equivalent electric current in the aperture and the integration is over the area of the antenna. If $w\leq \rmeas \pi/M$, cf.~\eqref{eq:refStep1}, then $\vec{E}_t$ will be approximately constant over the aperture of the \ac{MA} and $V$ will be proportional to the electric field strength. This is depicted in the lower part of Fig.~\ref{fig:probeWidth} and leads to
\begin{equation}\label{eq:dminResolution}
\rmeas\geq (R+\Delta R)\frac{w}{\lambda/2}.
\end{equation}
\eqref{eq:dminResolution} is a slightly stricter requirement than \eqref{eq:dminHpbw}. Therefore, \eqref{eq:dminResolution} is used hereafter. Note that \eqref{eq:dminResolution} can be applied for any circular cut if $R=\Rsph$, \ie\ the radius of the smallest sphere enclosing the \ac{EUT}. Moreover, it is noted that $\rmeas/(R+\Delta R)$ gives an upper bound to $w/(\lambda/2)$.

In the below examples \iac{SWE} truncated by using $\Delta R=\lambda$, see App.~\ref{app:swe}, is used for \nf\ calculations and~\cite{Delgado1999} is used for modeling of the \ac{SGH} aperture fields. The curvature of the field lines of $\vec{J}_a$ is neglected corresponding to long flared sections of the horns. The voltage is calibrated to get a correct \ac{TRP} value in the \ff\ of the \ac{EUT}.

\begin{figure}[]
\begin{center}
\includegraphics[width=\columnwidth]{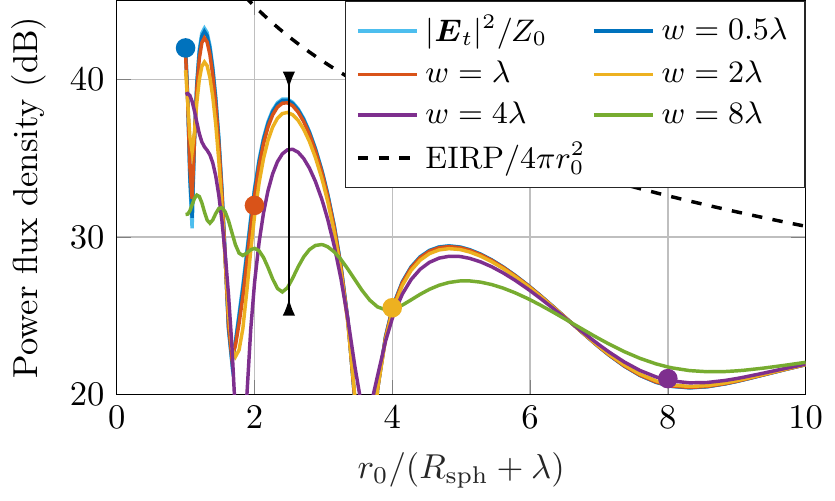}\\
\caption{A radial cut at $(\theta,\phi)=(\pi/2,0)$ measured with different \acp{MA} of width $w$. Note how the \nf\ power flux density deviates from the \ff\ (dashed black).}
\label{fig:radialCut}
\end{center}
\end{figure}

\begin{figure}[]
\begin{center}
\includegraphics[width=\columnwidth]{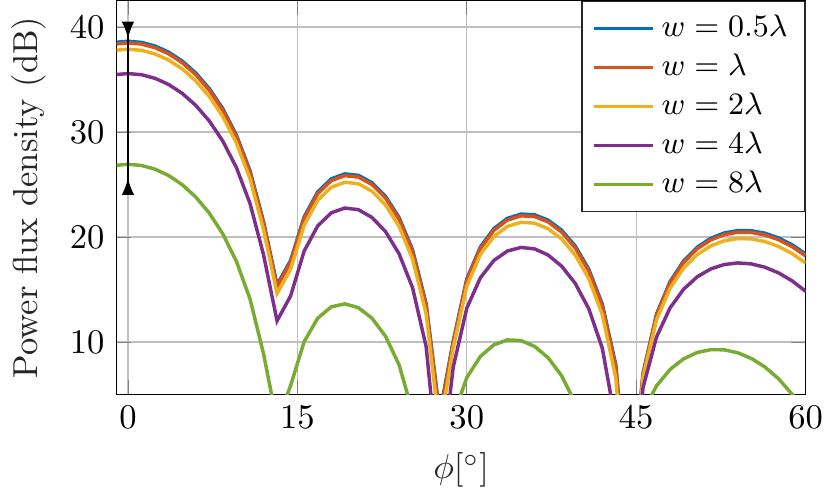}
\caption{Horizontal cut at $\theta=\pi/2$ at $\rmeas=2.5(\Rsph+\lambda)$. The maximum \ac{MA} width is $1.25\lambda$. Use of larger \acp{MA} leads to measurement errors.}
\label{fig:tangentialCut}
\end{center}
\end{figure}

An $8\times 8$ array of vertical half-wave dipoles at 28 GHz, width 4.3 cm and height 30 cm is used. The exaggerated height-to-width ratio is chosen to increase the deviation of the \nf\ cut from the \ff. In Figs~\ref{fig:radialCut} and \ref{fig:tangentialCut} probed power flux density using different widths $w$ is depicted. The radial cut of Fig.~\ref{fig:radialCut} is at the positive $x$-axis.  Minimum test distances~\eqref{eq:dminResolution} are indicated by dots, and the corresponding lines show small deviations at larger test distances. For the $8\lambda$ probe the minimum test distance is $16(R+\lambda)$. It is observed that a $w=0.5\lambda$ can be used for power flux density measurement as close to the \ac{EUT} as $\Rsph+\lambda$. Power flux density in the near field is not close to $\EIRP/4\pi\rmeas^2$ since \ac{EIRP} is a \ff\ quantity.  The horizontal cut of Fig.~\ref{fig:tangentialCut} is at $\rmeas =2.5(\Rsph+\lambda)$ and $\theta=\pi/2$. At this distance, $w\leq 1.25\lambda$ is suitable for measuring the power flux density. The intersection point of the cuts of Figs~\ref{fig:radialCut}-\ref{fig:tangentialCut} is depicted by vertical black lines.

\begin{figure}[]
\begin{center}
\includegraphics[width=\columnwidth]{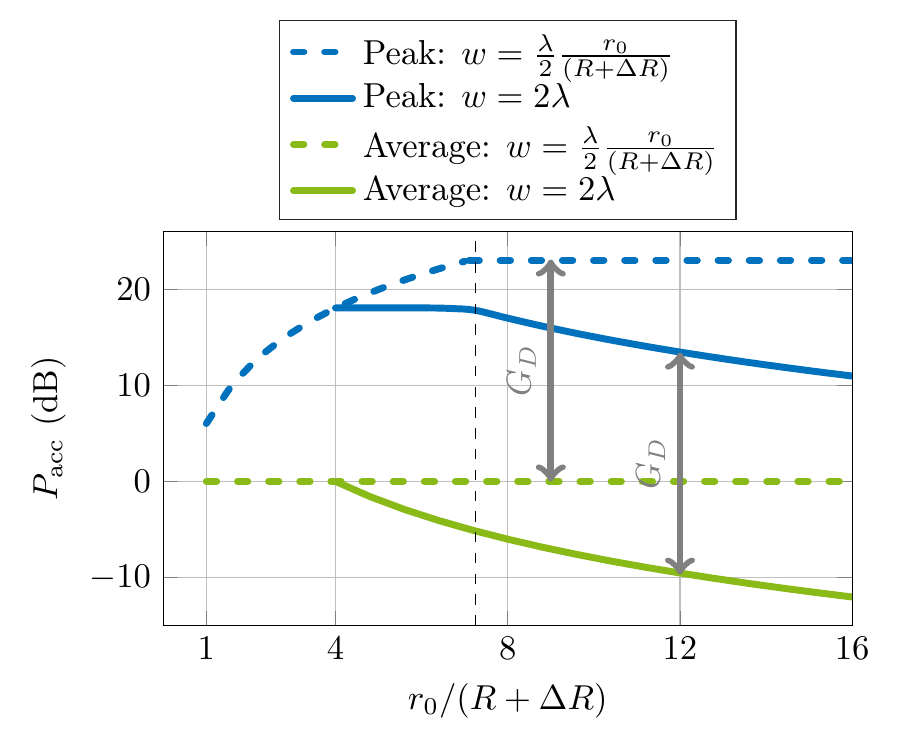}
\caption{Schematic link budget for an optimally large measurement antenna $w=(\lambda/2)\rmeas/(R+\Delta R)$, or a $w=2\lambda$ \ac{MA}. }
\label{fig:linkBudget}
\end{center}
\end{figure}

Finally, a link budget example for the accepted power $P_\text{acc}=A_\text{eff}S_r$ at the \ac{MA} port is given, see Fig.~\ref{fig:linkBudget}. The optimum width $w\propto\rmeas$ implies that the optimum effective area $A_\text{eff}^\text{opt}\propto\rmeas^2$. According to \eqref{eq:angularAverage} $\fsav{S_r} = \TRP/(4\pi \rmeas^2)$. Hence, the optimum $\fsav{P_\text{acc}}$ is constant (dashed green curve), \ie\ energy conservation. If a constant \ac{MA} is used, $\fsav{P_\text{acc}}$ is proportional to $1/\rmeas^2$  (solid green curve). As an example, the peak signal level is assumed constant near the \ac{EUT}, and beyond some breakpoint $\propto 1/\rmeas^2$. The peak accepted power, follows the same trend if a constant \ac{MA} is used and the minimum test distance is respected (solid blue curve). Using an optimum \ac{MA} yields $\max(P_\text{acc})$ proportional to  $\rmeas^2$ close to the antenna and constant beyond the breakpoint (dashed blue curve). Note that the peak to average ratio, the antenna directivity $G_D$ in the \ff\ region, is reduced in the \nf. The maximum \ac{SNR} is achieved when the probe is used at the minimum test distance~\cite{IEEE149-1979}. 

\subsection{Near-field impact on the TRP algorithm}
\begin{figure}[]
	\begin{center}
		\includegraphics[width=\columnwidth]{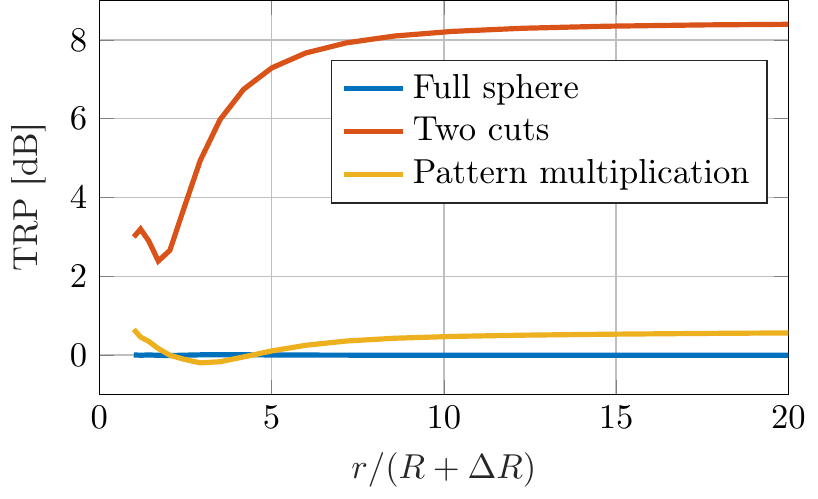}
		\caption{Comparison of the full sphere and two cuts grids, and pattern multiplication in the nearfield. A macro basestation antenna is used with fully correlated antenna element weights.}
		\label{fig:trpMethodNearfield}
	\end{center}
\end{figure}

Performance of some proposed \ac{TRP} methods applied in the \nf\ and for the antenna used in Fig.~\ref{fig:macro2TrpError} are shown in Fig.~\ref{fig:trpMethodNearfield}. The overestimation of the two cuts result is reduced at distances close to the \ac{EUT} and \ac{PM} performs well at all distances.

\section{Conclusions}

Challenges with \ac{OTA} measurement of \ac{TRP} in the \ac{MMW} region have been addressed. These challenges are the angular resolution and the search for worst case antenna configuration.

For uncorrelated emissions, $15^\circ$ sampling can be used at the expense of adding margins up to $2$ dB. Different grid types, two or three orthogonal cuts of full sphere, can be used with no need for alignment of the measurement grid to the \ac{EUT}. In the adjacent bands $uv$-plane pattern multiplication can be used on two-cut data to reduce uncertainty, both in \ff\ and \nf.

Secondly, a beam sweeping test signal is presented. Beam sweeping leads to wider beams for correlated emissions, \eg\ emissions at harmonics, and therefore a more relaxed angular sampling can be applied. Beam sweeping will also reduce the number of test configurations to a single one. Furthermore, it is closer to real use conditions for devices with beam-forming and tracking capabilities.

Additionally, it has been demonstrated that \nf\ measurements of radial power flux density can be used for \ac{TRP} assessment if the measurement antenna and test distance are selected by standard recommendations~\cite{IEEE149-1979}. Other sources of error, \eg\ alignment and field curvature effects, need further investigation.

Finally, for each finite test distance there is an optimal measurement antenna that will provide an average accepted power that is independent of test distance. Therefore, whether measurements need to be performed in the \nf\ is not fundamentally a question of path loss and signal levels, but rather a matter of measurement chamber size and equipment.

\appendices

\section{Sampling criteria}\label{app:sampling}

The resolution needed to correctly characterize an electric, or magnetic, field component in a circular cut is presented in this section. Without loss of generality the cut is defined as the set $\theta=\pi/2$ and $r=d$, \ie, a circle of radius $d$ in the $xy$-plane. Other circular cuts needed to cover a sphere are obtained by rotating the  \ac{EUT}. This field is periodic and can be expanded in the following series
\begin{equation}\label{eq:fieldModesIOnCircularCut}
E(d,\phi) \approx \sum\limits_{m=-\ceil{M}}^{\ceil{M}} a_m(d) \eu^{\ju m\phi},
\end{equation}
where $M = kR+N$~\cite[Eq.~(5.73)]{Hald1998} and $\ceil{\cdot}$ denotes rounding to nearest greater integer. A margin $N=10$ is commonly in use and corresponds to a dynamic range well above 40 dB in the measured data. Furthermore, $R$ is the radius of the smallest cylinder that encloses the \ac{EUT} and has the $z$-axis as symmetry axis. To demystify the subject of choosing $N$, $M$ is rewritten in terms of wave number and distances as
\begin{equation}\label{eq:truncationLimit}
M = kR+N = k(R+\Delta R).
\end{equation}
It is noted that the margin $\Delta R=(N/2\pi)\lambda$ can be interpreted as an added length resulting in an effective electrical length $R+\Delta R$ of the \ac{EUT}. The extra length $\Delta R$ is on the order of one wave length, cf.~\cite[Fig.~10]{Yaghjian1986} where $\Delta R=\lambda$ is used. The actual choice of $\Delta R$ or $N$ is related to the approximation error in Eq.~\eqref{eq:fieldModesIOnCircularCut}. To accurately determine the coefficients $a_m(d)$, and indirectly the field $E$, the angular sampling step needed is~\cite[Eq.~(5.76)]{Hald1998}

\begin{equation}\label{eq:refStep1}
\Delta\phi = \frac{\pi}{M} = \frac{\lambda/2}{R+\Delta R}.
\end{equation}
Note, $M$ is independent of $d$ and the circular cut can be taken anywhere in the radiating \nf\ or in the \ff\ region~\cite{Hald1998}. From the Parseval Theorem it follows that the angular average, which is the relevant quantity for \ac{TRP}, is
\begin{equation*}
\frac{1}{2\pi}\int_0^{2\pi}|E|^2(\phi) \mathrm{d}\phi \approx  \sum_{m=-\ceil{M}}^{\ceil{M}}|a_{m}|^2.
\end{equation*}
This suggests that the sampling needed for a correct average value is $\pi/M$. A major aim of this study is to investigate how densely angular measurements need to be taken in order to have a decent accuracy in the calculated \ac{TRP} value. For this purpose the reference angular step is defined by using $\Delta R=0$, resulting in
\begin{equation}\label{eq:refStep2}
\Delta\phi_\text{ref} = \frac{\pi}{kR}=\frac{\lambda/2}{R}.
\end{equation}
\begin{figure}
\begin{center}
\includegraphics[width=\columnwidth]{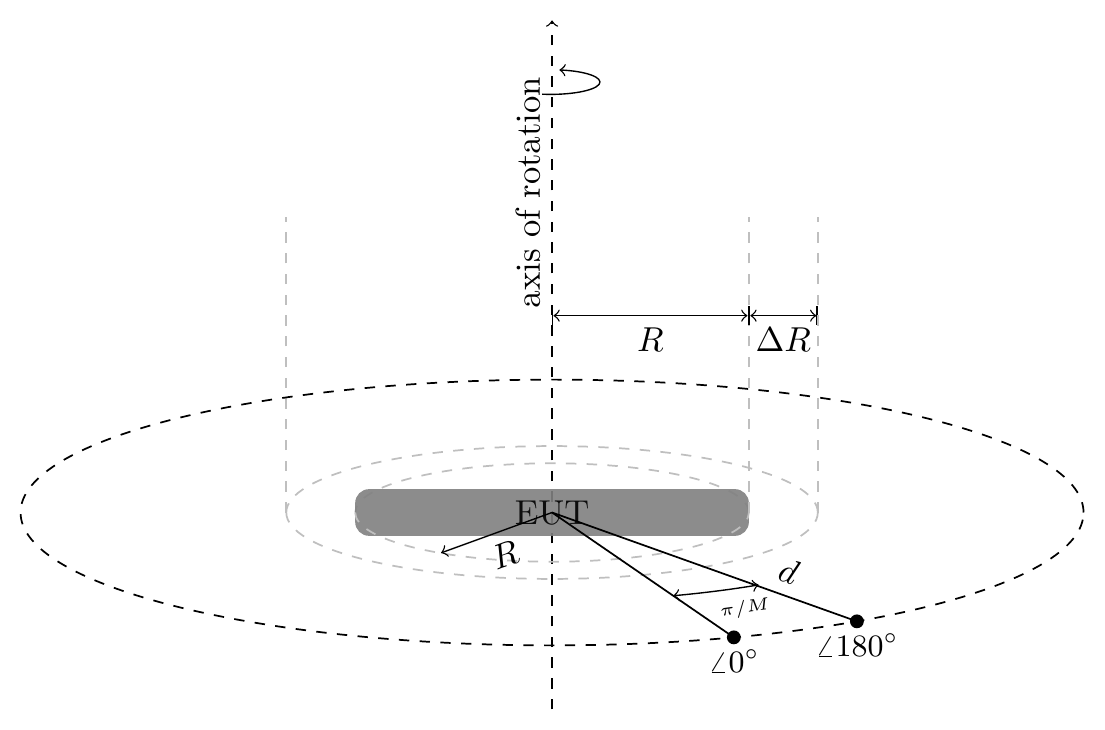}
\caption{The angular resolution for an angular cut of the \ac{EUT}. The angle $\pi/M$ denotes the angle over which the phase of the highest order modes $\exp{\pm\ju M\phi}$ changes phase by $180^\circ$.}
\label{fig:angularSampling}
\end{center}
\end{figure}
When the spherical coordinates~\eqref{eq:sphericalCoordinates} are used, the maximum radius in any constant $\phi$ cut is the radius $\Rsph$ of the smallest sphere enclosing the \ac{EUT}, and the effective radius for the $\phi$ cuts is the radius $\Rcyl$ of the smallest $z$-directed and $z$-axis centered circular cylinder that encloses the \ac{EUT}~\cite{Hald1998}. Hence, the reference angular steps are defined as
\begin{equation*}
\left\{
\begin{aligned}
\Delta\theta_\text{ref} &= \frac{\lambda/2}{R_\mathrm{sph}},\\
\Delta\phi_\text{ref} &= \frac{\lambda/2}{R_\mathrm{cyl}}
\end{aligned}
\right.
\end{equation*}

\section{Spherical wave expansion}\label{app:swe}
\begin{figure}[h]
\begin{center}
\includegraphics[width=\columnwidth]{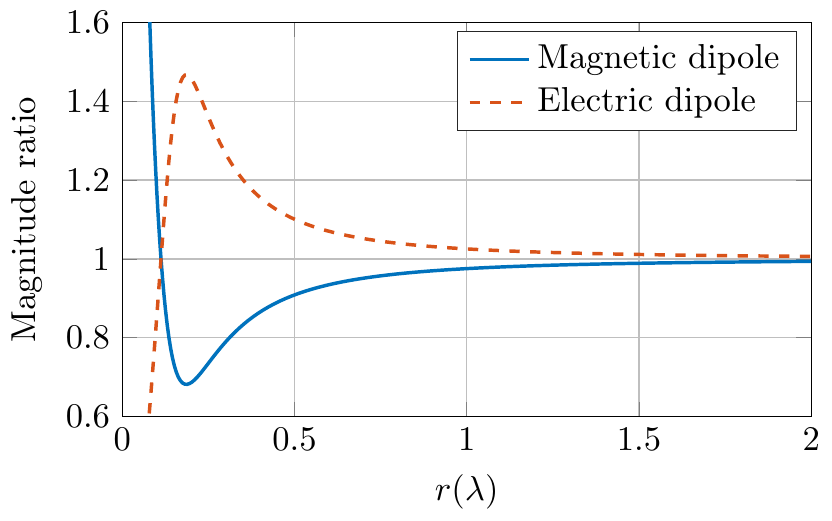}
\caption{Ratio of \ff\ and \nf\ radial power flux, $|\vec{E}_t|^2 Z_0^{-1}/\Re[\unitvec{r}\cdot(\vec{E}_t\times\vec{H}_t^*)]$ for infinitesimal magnetic and electric dipoles.}
\label{fig:smallDipoleApproximation}
\end{center}
\end{figure}

The electromagnetic fields of an antenna can be represented by a \acf{SWE}~\cite{Hald1998} as
\begin{equation}\label{eq:sweE}
\left\{
\begin{aligned}
\vec{E}_t(r,\theta,\phi)=&\sum\limits_{l=1}^\infty\sum\limits_{m=-l}^l a_{lm1}f_{l1}(kr)\vec{A}_{lm1}(\theta,\phi)\\&+a_{lm2}f_{l2}(kr)\vec{A}_{lm2}(\theta,\phi)\\
E_r(r,\theta,\phi)=&\sum\limits_{l=1}^\infty\sum\limits_{m=-l}^l a_{lm2}f_{l3}(kr)\unitvec{r}\cdot\vec{A}_{lm3}(\theta,\phi),
\end{aligned}
\right.
\end{equation}
and
\begin{equation}\label{eq:sweH}
\left\{
\begin{aligned}
\eta_0\vec{H}_t(r,\theta,\phi)&=\sum\limits_{l=1}^\infty\sum\limits_{m=-l}^l a_{lm1}f_{l2}(kr)\vec{A}_{lm2}(\theta,\phi)\\&-a_{lm2}f_{l1}(kr)\vec{A}_{lm1}(\theta,\phi)\\
\eta_0 H_r(r,\theta,\phi)=&\sum\limits_{l=1}^\infty\sum\limits_{m=-l}^l a_{lm1}f_{l3}(kr)\unitvec{r}\cdot\vec{A}_{lm3}(\theta,\phi)
\end{aligned}
\right.
\end{equation}
where $r\geq R$ and $R$ is the radius of the smallest sphere enclosing the antenna, hence the \ac{SWE} is valid both in the \nf\ and the \ff. The sums can be truncated at $l=kR+N$~\cite{Hald1998} where $k$ is the wave number, $R$ is the radius of the source, and $N$ is often chosen as 10. Furthermore, the radial functions are
\begin{equation*}
\left\{
\begin{aligned}
f_{l1}(kr)&=h_l^{(2)}(kr),\\
f_{l2}(kr)&=\frac{(kr h_l^{(2)}(kr))'}{kr},\\
f_{l3}(kr)&=\sqrt{l(l+1)}\frac{h_l^{(2)}(kr)}{kr}
\end{aligned}
\right.
\end{equation*}
where $h_l^{(2)}(kr)$ is the spherical Hankel function of the second kind, ensuring outgoing spherical waves. The angular functions are
\begin{equation*}
\left\{
\begin{aligned}
\vec{A}_{lm1}(\theta,\phi)&=\frac{1}{\sqrt{l(l+1)}}\nabla Y_{lm}(\theta,\phi)\times \vec{r},\\
\vec{A}_{lm2}(\theta,\phi)&=\frac{1}{\sqrt{l(l+1)}}r\nabla Y_{lm}(\theta,\phi),\\
\vec{A}_{lm3}(\theta,\phi)&=\unitvec{r}Y_{lm}(\theta,\phi),
\end{aligned}
\right.
\end{equation*}
where $Y_{lm}(\theta,\phi)$ are Spherical harmonics and $\nabla$ is the Nabla operator. Note the point-wise orthogonality between the tangential $n=1,2$ functions
\begin{equation}
\vec{A}_{lm2}(\theta,\phi)=\unitvec{r}\times\vec{A}_{lm1}(\theta,\phi).
\end{equation}
Asymptotically as $kr\rightarrow\infty$
\begin{equation*}
|f_{ln}(kr)| \rightarrow 
\begin{cases}
(kr)^{-1}&n=1,2\\
(kr)^{-2}&n=3\\
\end{cases}
\end{equation*}
Specifically for $l=1$, $m=0$
\begin{equation}
\left\{
\begin{aligned}
f_{11}(z) &=-\eu^{-\ju z}\frac{1}{z}-\eu^{-\ju z}\frac{\ju}{z^2}\\
f_{12}(z) &=-\eu^{-\ju z}\frac{\ju }{z}+\eu^{-\ju z}\frac{- z +\ju }{z^3}\\
\vec{A}_{101}&=\sqrt{\frac{3}{8\pi}}\sin\theta\unitvec{\phi}
\end{aligned}
\right.
\end{equation}
These expressions can be used to compare the \nf\ and \ff\ approximation of power density for the modes $lmn=101$ and $102$, \ie\ a magnetic and electric infinitesimal dipole, see Fig.~\ref{fig:smallDipoleApproximation}. For dipoles, it is seen that the \ff\ approximation, $r\rightarrow\infty$, is a good approximation as close as $1\lambda$ from the source. Since any antenna current can been represented as a super-position of dipole sources, this indicates that the approximation is probably valid also close to larger antennas. Note, when the form factor of \acp{EUT} is not close to a sphere, only a small fraction of measurement points will be close. Hence, the effect of the \ac{TRP} will be small.

The \ac{TRP} of the \ac{SWE} is calculated as
\begin{equation}\label{eq:trpFromSwe}
\ac{TRP} =  \sum\limits_{n=1}^2\sum\limits_{l=1}^\infty\sum\limits_{m=-l}^l |a_{lmn}|^2.
\end{equation}
The \ac{SWE} can be used to retreive \nf\ data from \ff\ data, \ie, back-propagation.
The following procedure is used: \FF\ data $r E_\theta(\theta,\phi)$ and $rE_\phi(\theta,\phi)$ is sampled on a full sphere. This data is used to calculate $a_{lmn}$. Typically 
$l\leq L=\ceil{k(\Rsph+\Delta R)}$ where the actual $\Delta R$ depends on the accuracy of the data and the desired precision. For evaluation at radius $r$, the truncation limit $L$ is reduced to $kr$ to control amplification of noise~\cite{Friden2003}. Electric and magnetic fields are then calculated by using \eqref{eq:sweE} and \eqref{eq:sweH}, respectively. The back-propagation error is defined as the change in \TRP\ caused by the adaptive truncation.


\section{Rotations}\label{app:rotations}

A rotation matrix can be written as~\cite{EulerRotation}
\begin{equation}
\mat{R}(\unitvec{n},\gamma)=\exp(\mat{A}\gamma)=\mat{I} + \mat{A}\sin\gamma + \mat{A}^2(1-\cos\gamma)
\end{equation}
where the generator matrix
\begin{equation}
\mat{A}=\begin{bmatrix}
0 &-n_z& n_y\\n_z&0&-n_x\\-n_y&n_x&0
\end{bmatrix},
\end{equation}
the rotation axis $\unitvec{n}=n_x\unitvec{x}+n_y\unitvec{y}+n_z\unitvec{z}$,
and $\gamma$ is the rotation angle in the positive sense around the rotation axis. Note that $\mat{A}^3=-\mat{A}$ can be used to reduce the infinite Taylor series and identify the cosine and sine functions.

The far-field pattern for rotated point sources reads
\begin{equation}\label{eq:arrayFactorRotated}
\vec{F}(\theta,\phi) = \sum\limits_n w_n\eu^{\ju k\unitvec{r}(\theta,\phi)\cdot\mat{R}(\unitvec{n},\gamma)\cdot\vec{d}_n}.
\end{equation}

To generate random rotations the rotation axis is parametrized using spherical coordinates, see~\eqref{eq:sphericalCoordinates},
\begin{equation}
\unitvec{n}(\alpha,\beta) = \unitvec{r}(\alpha,\beta),
\end{equation}
and $\alpha\in[0,\pi/2]$, $\beta,\gamma\in[-\pi,\pi]$ are used to generate random rotations. Note that the rotation axis can be restricted to the upper hemi-sphere since $\mat{R}(\unitvec{n},\gamma)=\mat{R}(-\unitvec{n},-\gamma)$.

\section{The $uv$-plane integration}\label{app:integration}
As described in Sec.~\ref{sec:sparsity} the integration of eq.~\eqref{eq:uvIntegration} involves a singularity along the edge of the visible region in $uv$ coordinate system, \ie, $u^2+v^2=1$. In order to resolve this singularity, the following change of variables is used
\begin{equation*}
\left\{
\begin{aligned}
u = \sqrt{1-\xi^2}\cos \alpha,\\
v = \sqrt{1-\xi^2}\sin \alpha.
\end{aligned}
\right.
\end{equation*}
The infinitesimal solid angle is
\begin{equation*}
\mathrm{d}\Omega=\frac{\mathrm{d}u\mathrm{d}v}{\sqrt{1-u^2-v^2}}=\mathrm{d}\xi\mathrm{d}\alpha.
\end{equation*}
Therefore, the radiated power in the forward hemisphere becomes
\begin{equation}
\begin{aligned}
P_\text{rad}^\text{fwd} = &\iint_{u^2+v^2\leq 1} S^\text{fwd}_r(u,v)  r^2 \dOmega\\
= & r^2\int_{\alpha=0}^{2\pi}\int_{\xi=0}^{1}S^\text{fwd}_r(\xi,\alpha)\mathrm{d}\xi\mathrm{d}\alpha,
\end{aligned}
\end{equation}
where $S^\text{fwd}_r$ is the power flux density in the forward hemisphere. Finally, two test cases are given for code testing, see Table~\ref{tab:testCases}. Test case (a) is separable in $uv$-coordinates, and therefore the \ac{PM} estimate is identical to the true \ac{TRP}. 
\begin{table}
\begin{center}
\caption{Two test cases for two-cuts and pattern multiplication estimates of $\TRP$. }
\label{tab:testCases}
\begin{tabular}{rcc}
&(a)&(b)\\\toprule
$S_r(\theta,\phi)$ @ $r=1$&$\sin^2\theta$&$\sin^2\theta\cos^2\phi$\\
$S_r(u,v)$ &$1-v^2 $&$1-u^2-v^2 $\\
$S_r(u,v=0)$&1&$1-u^2$\\
$S_r(u=0,v)$&$1-v^2$&$1-v^2$\\
$\TRP_\mathrm{2cuts}$&$3\pi$&$2\pi$\\
$\TRP_\mathrm{3cuts}$&$8\pi/3$&$4\pi/3$\\
$\TRP_\mathrm{PM}$&$8\pi/3 $&$8\pi/5$\\
$\TRP$&$8\pi/3$&$4\pi/3$\\
\bottomrule
\end{tabular}

\end{center}
\end{table}


\section*{Acknowledgment}

The authors would like to thank the ANSI C63.26 working group, and Ericsson colleagues for fruitful discussions. 

\ifCLASSOPTIONcaptionsoff
  \newpage
\fi


\bibliographystyle{IEEEtran}
\bibliography{IEEEabrv,refs}

\begin{IEEEbiography}[{\includegraphics[width=2.5cm]{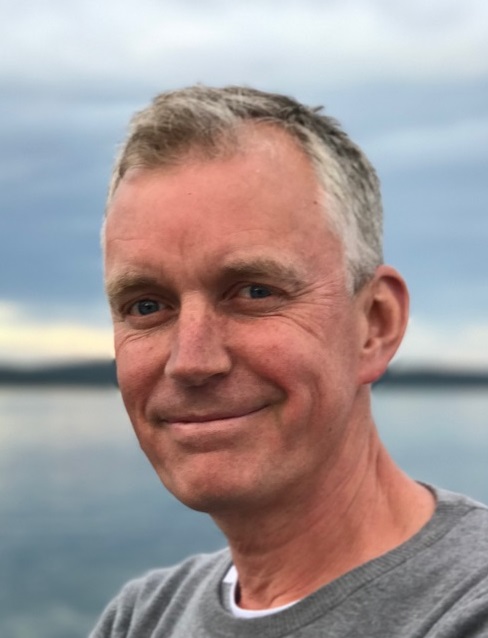}}]{Jonas Frid\'en}
was born in 1965. He received the B.S. degree in mathematics and physics and the Ph.D. degree in theoretical physics from the University of G\"{o}teborg, G\"{o}teborg, Sweden, in 1987 and 1996, respectively. Since 2002, he has been with Ericsson Research, Ericsson AB, G\"{o}teborg, Sweden. In 1996 --1999, he was a lecturer with the College University of Bor\r{a}s. In 1999, he was with Ericsson Microwave Systems AB, where he worked with radar antennas, radar system, and radome design. He has also been a member of the European Electromagnetic Data Interface Group. His major areas of research are electromagnetic compliance, OTA measurement techniques, near field retrieval techniques, antenna theory, bandwidth limitations of antennas, and MIMO antenna system performance.
\end{IEEEbiography}
\begin{IEEEbiography}[{\includegraphics[width=2.5cm]{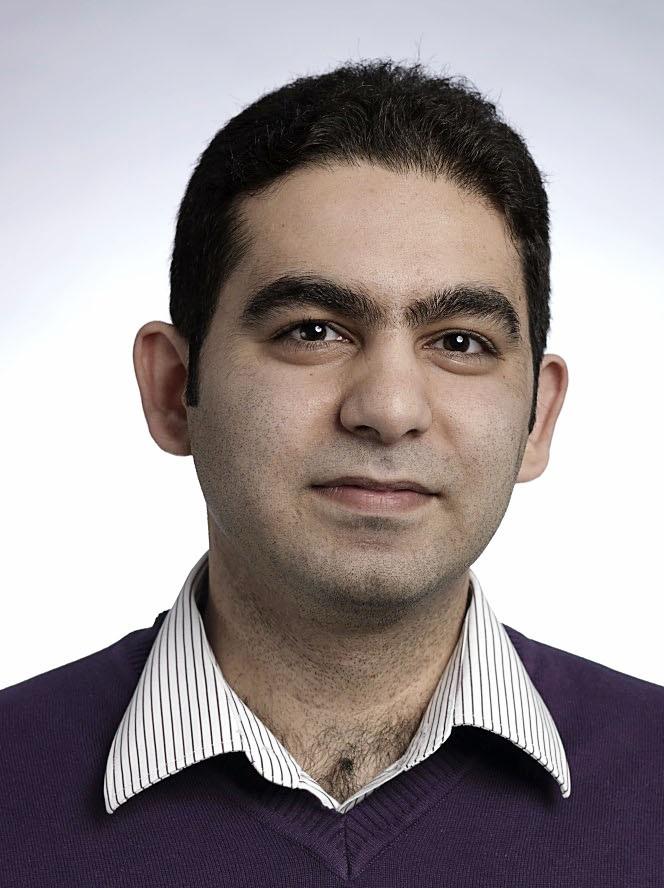}}]{Aidin Razavi}
was born in 1982. He received the M.S. degree in microwave engineering from Tarbiat Modares University, Tehran, Iran, in 2007 and the Ph.D. degree from Chalmers University of Technology, Gothenburg, Sweden, in 2016. Between 2007 and 2009, he was with HT Telecom Co., Iran as senior engineer, and from 2009 to 2011 with Huawei Technologies Co., as radio network planning and optimization engineer. Since 2017, he has been with Ericsson Research, Ericsson AB, G\"{o}teborg, Sweden. His major areas of research include antenna theory, OTA measurements, MIMO antenna system performance, and optimal antenna apertures.
\end{IEEEbiography}
\begin{IEEEbiography}[{\includegraphics[width=2.5cm]{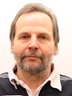}}]{Anders Stjernman}
was born in 1956. He received the M.S. degree in Physical Engeering from Lund University 1982 and Ph.D. degree in Space Physics from Ume\r{a} University 1995. Since 2002 he has been with Ericsson Research, Ericsson AB, G\"{o}teborg, Sweden.
In 1995--2002, he was with Ericsson Microwave Systems AB, where he worked with radar antennas, radar system,  radome design  and bluetooth antennas.  In 1986--1995 he worked at Swedish Institute of Space Physics (IRF) in Kiruna with radio remote sensing. His major areas of research are antenna theory, bandwidth limitations of antennas, and MIMO antenna system performance.
\end{IEEEbiography}

\vfill

\end{document}